\documentclass[aps,superscriptaddress,prd,eprint,nofootinbib,showkeys]{revtex4-2}
\pdfoutput=1
\usepackage{graphicx}
\usepackage{dcolumn}
\usepackage{bm}
\usepackage{amsmath,amsfonts,amsthm,amssymb}
\usepackage{graphics}
\usepackage{cancel}
\usepackage{multirow}
\usepackage{caption}
\usepackage{subcaption}
\usepackage{color}
\usepackage[normalem]{ulem}

\usepackage{amssymb}

\usepackage{graphicx}
\usepackage{amsmath}
\usepackage{bbold}
\usepackage{yfonts}
\usepackage[colorlinks=true,linktocpage=true,linkcolor=blue,citecolor=blue]{hyperref}
\usepackage{placeins}
\usepackage{bm} 
\usepackage{nicefrac}
\usepackage{slashed}
\usepackage{marginnote}

\begin{document}

\preprint{APS/123-QED}

\title{ Particle diffusion in magnetized QCD matter with a realistic equation of state} 

\author{Arnab Sarkar}
\email{24dr0274@iitism.ac.in}
\affiliation{Department of Physics, Indian Institute of Technology (Indian School of Mines) Dhanbad, Jharkhand 826004, India}

\author{Manu Kurian}
\email{manukurian@iitism.ac.in}
\affiliation{Department of Physics, Indian Institute of Technology (Indian School of Mines) Dhanbad, Jharkhand 826004, India}
\begin{abstract}
   In this study, we investigate particle number diffusion in an interacting hot QCD medium in the presence of an external magnetic field. Starting from an effective covariant description of the medium in terms of quasiparticle degrees of freedom, we employ a Chapman–Enskog-like iterative expansion of the distribution function to derive the second-order relativistic magnetohydrodynamic evolution equation for the number diffusion current. The magnetic field introduces additional tensorial structures into the diffusion dynamics, thereby modifying the transport properties of the medium. We further obtain the corresponding Navier–Stokes limit and examine the influence of medium interactions and the magnetic field on the diffusion current and its associated transport coefficients. Our results show that medium interactions suppress the magnetic field-induced anisotropy in particle diffusion. This study provides a consistent framework for investigating dissipative particle transport in magnetized, interacting QCD matter and may serve as an important ingredient in developing realistic hydrodynamic descriptions of heavy-ion collisions.
\end{abstract}
\maketitle
\section{Introduction} \label{Intro}
Heavy-ion collision experiments at the Relativistic Heavy Ion Collider (RHIC) and the Large Hadron Collider (LHC) provide a unique opportunity to explore strongly interacting matter at extreme conditions of high temperature and density. These high-energy collisions create a short-lived state of Quantum Chromodynamics (QCD) matter known as the Quark-Gluon Plasma (QGP), which exhibits collective behavior that is remarkably well described by relativistic hydrodynamics~\cite{Gale:2013da, Heinz:2013th}. In particular, hydrodynamic framework has successfully reproduced a range of experimental observables, including the transverse-momentum spectra and anisotropic flow of the produced particles, demonstrating that the QGP exhibits a fluid-like behavior~\cite{Romatschke:2007mq, Schenke:2010rr}. This close connection between experimental observations and hydrodynamic descriptions has made relativistic fluid dynamics an important framework for investigating the microscopic properties of QGP. Intense magnetic fields are generated during heavy-ion collision events, particularly during their early stages~\cite{Deng:2012pc, Bzdak:2011yy}. Though these fields are expected to decay fast as the system evolves, its subsequent evolution within the QCD medium can be more complex. This is because the time-varying magnetic field can induce electric currents in the conducting QGP, which may, in turn, slow down the decay of the field and sustain its presence for a longer period than initially expected. The lifetime and evolution of the magnetic field in the QCD medium are still not completely known. Experimental observations of the heavy mesons directed flow and the splitting between the directed flows of $D^0$ and $\bar{D}^0$ mesons at the RHIC and LHC~\cite{ALICE:2019sgg, STAR:2019clv} provide phenomenological constraints on the strength and evolution of the magnetic field~\cite{Das:2016cwd, Jiang:2022uoe}. Considerable theoretical studies have been devoted to understanding the impact of magnetic fields on the QCD thermodynamics, its transport properties, heavy quark dynamics, and dilepton spectra~\cite{Karmakar:2019tdp, Kurian:2017yxj, Ghosh:2024fkg, Ghosh:2022vjp, Goswami:2023eol, Rath:2023abm, Ghosh:2023ghi, Gangopadhyaya:2022olm, Satapathy:2021cjp, Kurian:2019nna, Tuchin:2013ie, Hattori:2020htm, Kurian:2018qwb, Wang:2022jxx, Panda:2025yxw}.  Furthermore, the magnetic field combined with chiral anomaly in QCD medium can lead to anomalous transport phenomena, including the Chiral Magnetic Effect~\cite{Fukushima:2008xe, Hattori:2016emy, Kharzeev:2015znc, Shi:2019wzi}.

Relativistic hydrodynamics needs to be coupled with electromagnetic fields to describe the evolution of magnetized QCD matter~\cite{Inghirami:2019mkc, Denicol:2019iyh,  Hattori:2022hyo}. Since the QGP is a strongly interacting medium with finite dissipative effects, its evolution cannot be described by ideal hydrodynamics alone. The evolution of the shear-stress tensor, bulk viscous pressure, and the particle number diffusion current are required for a consistent description of the medium at a non-vanishing chemical potential. The evolution of the shear-stress tensor and bulk viscous pressure has been extensively studied with and without a magnetic field~\cite{DenicolTau, Florkowski:2015lra, Bhadury:2020ivo, Hattori:2017qih, Singh:2024leo}. However, the dynamics of the dissipative charge current, especially in the presence of electromagnetic fields, has received comparatively less attention. The presence of a magnetic field can substantially modify the dissipative transport properties of the QGP. In particular, the magnetic field introduces a preferred spatial direction, making the medium response anisotropic and leading to multiple independent transport coefficients depending on the strength and orientation of the field~\cite{Li:2017tgi, Modak:2026pmi, Rath:2025zhw, Shaikh:2022sky}. Some attempts have been made to understand the charge diffusion in the magnetized QCD medium~\cite{Dash:2020vxk, Denicol:2018rbw, Panda:2021pvq, Das:2019ppb, Dash:2022xkz}. A realistic description of QGP dynamics within dissipative relativistic hydrodynamics requires the thermodynamic properties of QCD and its microscopic transport behavior to be incorporated consistently. These properties enter the hydrodynamic evolution through the equation of state (EoS) and the corresponding transport coefficients. In a kinetic-theory description, the effects of the non-ideal QCD EoS can be accounted for by treating quarks, anti-quarks, and gluons as effective quasiparticle degrees of freedom, whose properties are constrained to reproduce the thermodynamics of QCD~\cite{Bluhm:2006yh, Bannur:2006js, Mykhaylova:2019wci, Chandra:2011en, Plumari:2011mk, Berrehrah:2016vzw}. A recent study~\cite{Daher:2024vxk} has analyzed the second-order evolution equation of the diffusion current with a realistic EoS. Extending this formulation to a magnetized QCD medium introduces additional magnetic field-dependent contributions to the diffusion evolution and modifies the associated transport coefficients. A systematic investigation of the interplay between the realistic EoS and magnetic field modified diffusion dynamics is therefore essential for developing a consistent second-order dissipative framework for magnetized QCD matter at finite chemical potential. This provides the motivation for the present study.

In the present study, we investigate the non-resistive second-order evolution equation of particle number diffusion while incorporating the realistic EoS effects for a magnetized QCD medium. The QCD EoS is incorporated within a quasiparticle description, in which the medium is represented in terms of effective quark, anti-quark, and gluon degrees of freedom. 
Thermodynamic consistency is maintained by introducing an additional mean-field contribution to the energy-momentum tensor. This mean-field contribution is also consistently incorporated into the Boltzmann equation. In this way, the effects of the non-ideal QCD EoS can be incorporated into the kinetic description while maintaining consistency between the thermodynamic and microscopic descriptions of the medium. We solve the effective Boltzmann equation within the relaxation-time approximation (RTA) using a Chapman–Enskog-like gradient expansion. We derive the second-order evolution equation for the particle-number diffusion current and investigate the combined effects of the magnetic field and the EoS on its evolution. Furthermore, in the Navier–Stokes limit, we obtain the diffusion current and demonstrate how the magnetic field modifies the corresponding anisotropic diffusion coefficients for a system with quasiparticle degrees of freedom.
 
The manuscript is organized as follows. In Sec.~\ref{Qp-MHD}, we derive the magnetohydrodynamic evolution equation of particle number current in a magnetized QCD medium within an effective kinetic theory. In Sec.~\ref{Result}, we discuss the interplay of magnetic field and non-ideal EoS on the particle transport in QCD medium. Finally, Sec.~\ref{Conc} summarizes the main findings and concludes the manuscript.

{\it Notation and Conventions}:  We adopt natural units throughout the analysis, such that $c=k_B=\hbar=1$ and and use the metric convention $g_{\mu\nu}={\text {diag}}\,(1,-1,-1,-1)$. The four-velocity of the fluid is denoted by $u_\mu$ and is normalized according to $u^\mu u_\mu=1$. In the local rest frame (LRF) of the fluid, it takes the form $u^{\mu}=(1,0,0,0)$. The tensor projecting any four-vector onto the three-dimensional subspace orthogonal to the fluid velocity is defined as $\Delta^{\mu\nu} = g^{\mu\nu} - u^\mu u^\nu$, which satisfies $\Delta^{\mu\nu} u_\mu=0$. For projecting a rank-two tensor onto its symmetric, traceless, and $u^\mu$ orthogonal part, we introduce the rank-four projection tensor $\Delta^{\mu\nu}_{\alpha\beta}\equiv\frac{1}{2}(\Delta^\mu_\alpha\Delta^\nu_\beta +\Delta^\mu_\beta\Delta^\nu_\alpha)-\frac{1}{3}\Delta^{\mu\nu}\Delta_{\alpha\beta}$. One can define $\nabla^\mu u^\nu$ in terms of symmetric traceless tensor $\sigma^{\mu\nu}$, anti-symmetric vorticity tensor $\omega^{\mu\nu}$, and expansion scalar $\theta$ as $\nabla^\mu u^\nu = \sigma^{\mu\nu} + \omega^{\mu\nu} + \frac{1}{3} \Delta^{\mu\nu} \theta$, where $\sigma^{\mu\nu} = \frac{1}{2} (\nabla^\mu u^\nu + \nabla^\nu u^\mu) - \frac{1}{3} \Delta^{\mu\nu} \theta$, $\omega^{\mu\nu} = \frac{1}{2} (\nabla^\mu u^\nu - \nabla^\nu u^\mu)$, and $\theta = \partial_\mu u^\mu$.
\section{Particle Number Current in a Magnetized QCD Medium} \label{Qp-MHD}
\subsection{Equations of Motion}
A convenient description of electromagnetic interactions in relativistic matter is provided by the field strength tensor. 
The electromagnetic tensor $F^{\mu\nu}$ and its dual counterpart $\widetilde{F}^{\mu\nu}$ can be covariantly written as,
\begin{align}
    F^{\mu\nu} = E^\mu u^\nu - E^\nu u^\mu + \epsilon^{\mu\nu\alpha\beta} u_\alpha B_\beta,
    \\
    \widetilde{F}^{\mu\nu} = B^\mu u^\nu - B^\nu u^\mu - \epsilon^{\mu\nu\alpha\beta} u_\alpha E_\beta,
\end{align}
where $\epsilon^{\mu\nu\alpha\beta}$ is the Levi-Civita tensor. The electric and magnetic four-vectors are defined as $E^\mu = F^{\mu\nu} u_\nu$ and $B^\mu = \Tilde{F}^{\mu\nu} u_\nu$, respectively. In the local rest frame of the fluid, these four-vectors reduce to $E^\mu = (0,\textbf{E})$, $B^\mu = (0,\textbf{B})$ and $u^\mu = (1,\textbf{0})$, where $E^\mu$, $B^\mu$ are orthogonal to the fluid velocity $u^\mu$, $i.e.$, $E^\mu u_\mu = B^\mu u_\mu = 0$. In this covariant formulation, Maxwell's equations take the form,
\begin{align} \label{eq6}
   & \partial_\mu F^{\mu\nu} = \mathcal{J}^\nu,  
    &&\partial_\mu \widetilde{F}^{\mu\nu} = 0,
\end{align}
with $\mathcal{J}^\mu$ as the total electric charge four-current, which acts as the source of the electromagnetic field. The four-current can be expressed as the sum of conduction and diffusion contributions~\cite{Panda:2020zhr},
\begin{equation}\label{eq7}
    \mathcal{J}^\mu = j^\mu + d^\mu.
\end{equation}
Here, $d^\mu$ is the diffusion current and $j^\mu = n_q u^\mu$ is the conduction current with $n_q = q n_f$, where $n_f$ is the number density and $q$ is the electric charge. Solving Eq.~(\ref{eq6}) provides the evolution of the electromagnetic field. The relation between the electric field and the conduction current is described by Ohm's law, which can be expressed covariantly as, $j^\mu = \varsigma^{\mu\nu} E_\nu$, with $\varsigma^{\mu\nu}$ as the conductivity tensor. For an isotropic medium the conductivity tensor reduces to $\varsigma^{\mu\nu} \rightarrow \varsigma g^{\mu\nu}$, and Ohm's law consequently takes the form $j^\mu = \varsigma E^\mu$. In the present study, we consider the non-resistive limit, corresponding to an infinitely large electrical conductivity, $\varsigma \rightarrow \infty$. Since the conduction current induced by the partons remains finite, this limit leads to a vanishingly small electric field, $E^\mu \rightarrow 0$. Consequently, the electromagnetic field tensor reduces to the following form:
\begin{equation}
    F^{\mu\nu} \rightarrow B^{\mu\nu} = \epsilon^{\mu\nu\alpha\beta} u_\alpha B_\beta = -B b^{\mu\nu}, \label{B_munu}
\end{equation}
where $B^\mu B_\mu = -B^2$ and $b^\mu = \frac{B^\mu}{B}$ is the normalized magnetic four-vector satisfying, $b^\mu b_\mu = -1$. One can define an projection operator orthogonal to both $u^\mu$ and $b^\mu$ in presence of magnetic field \cite{Singh:2024leo},
\begin{equation}
    \Xi^{\mu\nu} = \Delta^{\mu\nu} + b^\mu b^\nu .
\end{equation}
The anti-symmetric tensor $b^{\mu\nu}$ is defined as, $b^{\mu\nu} = \frac{B^{\mu\nu}}{B}$. It is orthogonal to both $b^\mu$ and $u^\mu$, $i.e.$, $b^{\mu\nu} b_\nu = b^{\mu\nu} b_\nu = 0$ and obeys the normalization condition, $b^{\mu\nu} b_{\mu\nu} = 2$. Furthermore, using the identities of the Levi-Civita tensor, one obtains, $b^{\mu\alpha} b_{\mu\beta} = \Xi^\alpha_\beta$. Using Maxwell's equations, the evolution equation for the magnetic field can be expressed as~\cite{Panda:2020zhr, Singh:2024leo},
\begin{equation}
    \epsilon^{\mu\nu\alpha\beta} \left(u_\alpha \partial_\mu B_\beta + B_\beta \partial_\mu u_\alpha \right) = \mathcal{J}^\nu .
\end{equation}
The energy-momentum tensor of the electromagnetic field, denoted by $\mathcal{T}^{\mu\nu}_{EM}$, takes the following form in the non-resistive limit:
\begin{equation}
    \mathcal{T}^{\mu\nu}_{EM} \rightarrow \mathcal{T}^{\mu\nu}_B = F^{\mu\lambda} F^\nu_\lambda + \frac{1}{4} g^{\mu\nu} F^{\alpha\beta} F_{\alpha\beta} = \frac{B^2}{2}(u^\mu u^\nu - \Delta^{\mu\nu} - 2 b^\mu b^\nu) .
\end{equation}
The evolution of the electromagnetic field can be related to its energy-momentum tensor through the following equation,
\begin{equation}
    \partial_\mu \mathcal{T}^{\mu\nu}_{EM} = -F^{\nu\lambda} \mathcal{J}_\lambda. \label{Evol_Tem}
\end{equation}
In the absence of external source currents, the total electric four-current consists solely of the contribution from the fluid, $i.e.$, $\mathcal{J}^\mu = \mathcal{J}^\mu_f$, where $\mathcal{J}^\mu_f = q \mathcal{N}^\mu$ and $\mathcal{N}^\mu$ denotes the particle number four-current. However, in heavy-ion collisions, spectator protons can act as external sources of electromagnetic fields and induce additional charge currents. In such situations, the total electric four-current can be decomposed into fluid and external contributions as,
\begin{equation}
    \mathcal{J}^\mu = \mathcal{J}^\mu_f + \mathcal{J}^\mu_{ext} . \label{total_J}
\end{equation}
The fluid electric four-current satisfies the charge conservation equation:
\begin{equation}
    \partial_\mu \mathcal{J}^\mu_f = 0 .
\end{equation}
In the presence of external currents, the conservation equation for the combined fluid and electromagnetic energy-momentum tensor acquires a source term. The total energy-momentum tensor is defined as,
\begin{equation}
    \mathcal{T}^{\mu\nu} = \mathcal{T}_f^{\mu\nu} + \mathcal{T}_{EM}^{\mu\nu},
\end{equation}
where $\mathcal{T}^{\mu\nu}_f$ and $\mathcal{T}^{\mu\nu}_{EM}$ denote the energy-momentum tensors of the fluid and electromagnetic field, respectively. Its conservation equation in the presence of an external current reads,
\begin{equation}
    \partial_\mu \mathcal{T}^{\mu\nu} = -F^{\nu\lambda} \mathcal{J}_{ext, \lambda}. \label{Evol_Ttot}
\end{equation}
The fluid evolution equation in the presence of an external electromagnetic field can be obtained using Eq.~(\ref{Evol_Tem}), Eq.~(\ref{total_J}) and Eq.~(\ref{Evol_Ttot}), yielding
\begin{equation}
    \partial_\mu \mathcal{T}^{\mu\nu}_f = F^{\nu\lambda} \mathcal{J}_{f, \lambda} . \label{Evol_Tmunu}
\end{equation}
\subsection{QCD Medium with Realistic EoS}
To describe a realistic QCD medium, the interactions among quarks and gluons must be taken into account. Within the kinetic theory framework, the effects of thermal medium interactions can be incorporated through temperature- and chemical-potential-dependent effective masses of quarks and gluons. However, introducing effective masses modifies the thermodynamic relations and can lead to thermodynamic inconsistencies~\cite{0608:Bannur}. To ensure thermodynamic consistency, an additional bag pressure term is introduced into the energy density and thermal pressure. Accordingly, the energy-momentum tensor and particle-number four-current of a viscous fluid with medium interactions can be expressed covariantly as~\cite{Mukherjee:2025dqp, 0608:Bannur},
\begin{align}
    \mathcal{T}^{\mu\nu}_f &= \mathcal{E}_0 u^\mu u^\nu - (\mathcal{P}_0 + \Pi) \Delta^{\mu\nu} + \mathcal{B} g^{\mu\nu} + \pi^{\mu\nu}, \label{T_munu}
    \\
    \mathcal{N}^\mu_f &= n_f u^\mu + V^\mu \label{N_mu},
\end{align}
where $n_f$ is net particle number density, $\mathcal{E}_0$ is the energy density, $\mathcal{P}_0$ is the thermal pressure, and $\mathcal{B}$ is bag pressure term. The dissipative terms are characterized by the shear stress tensor $\pi^{\mu\nu}$, bulk pressure $\Pi$, and number diffusion current $V^\mu$. The energy-momentum tensor can be written concisely as~\cite{Daher:2024vxk},
\begin{equation} \label{F_T_munu}
    \mathcal{T}^{\mu\nu}_f = \mathcal{E} u^\mu u^\nu - (\mathcal{P} + \Pi)  \Delta^{\mu\nu} + \pi^{\mu\nu},
\end{equation}
where $\mathcal{E}$ and $\mathcal{P}$ denote the thermodynamically consistent energy density and thermal pressure, respectively, defined as  $\mathcal{E} = \mathcal{E}_0 + \mathcal{B}$ and $\mathcal{P} = \mathcal{P}_0 - \mathcal{B}$. Using the appropriate projection operators, these thermodynamic quantities can be extracted from the equilibrium energy-momentum tensor and particle number four-current as,
\begin{align}
   & \mathcal{E} = u_\mu u_\nu \mathcal{T}^{\mu\nu}_f, && \mathcal{P} = -\frac{1}{3} \Delta_{\mu\nu} \mathcal{T}^{\mu\nu}_f, &&
    n_f = u_\mu \mathcal{N}^\mu .
\end{align}
Using grand canonical thermodynamic relations, the EoS can be expressed as,
\begin{align}
    \frac{\partial \mathcal{P}}{\partial \mu} = n_f, && \frac{\partial \mathcal{P}}{\partial T} = \frac{\mathcal{E} + \mathcal{P} - \mu n_f}{T}. \label{EoS}
\end{align}
For a non-magnetizable and non-polarizable medium interacting with an external magnetic field, the total energy-momentum tensor of the magnetohydrodynamic system, comprising the fluid and electromagnetic contributions, takes the form as follows,
\begin{equation}
    \mathcal{T}^{\mu\nu} = \mathcal{T}^{\mu\nu}_f + \mathcal{T}^{\mu\nu}_{EM} = \left(\mathcal{E} + \frac{B^2}{2} \right) u^\mu u^\nu - \left(\mathcal{P} + \frac{B^2}{2} + \Pi \right)\Delta^{\mu\nu} - B^2 b^\mu b^\nu + \pi^{\mu\nu} . 
\end{equation}
The fluid evolution equations follow from the sourced energy-momentum balance equation and the conservation of the particle number four-current. By projecting the energy-momentum conservation equation with fluid four velocity $u^\mu$ and and onto the subspace orthogonal to $u^\mu$, and using the conservation of four-number current we obtain~\cite{Panda:2020zhr, Singh:2024leo},
\begin{align}
    \partial_\mu \mathcal{N}^\mu_f &= 0,
    \\
    u_\nu \partial_\mu \mathcal{T}^{\mu\nu}_f &= 0,
    \\
    \Delta^\alpha_\nu \partial_\mu \mathcal{T}^{\mu\nu}_f &= -q B b^{\alpha\lambda} V_\lambda .
\end{align}
Substituting Eqs.~(\ref{F_T_munu}) and (\ref{N_mu}) into the above conservation equations leads to the following evolution equations:
\begin{align}
    \dot{n}_f + n_f \theta + \partial_\mu V^\mu &= 0 ,
    \\
    \mathcal{\dot{E}} + (\mathcal{E} + \mathcal{P} + \Pi)\theta - \pi^{\mu\nu}\sigma_{\mu\nu} &= 0 ,
    \\
    (\mathcal{E} + \mathcal{P} + \Pi) \dot{u}^\alpha - \nabla^\alpha(\mathcal{P} + \Pi) + \Delta^{\alpha}_\nu \partial_\mu \pi^{\mu\nu} &= - q B b^{\alpha\lambda} V_\lambda .
\end{align}
Here, the dot denotes the co-moving time derivative, $\dot{A}=u^\mu\partial_\mu A$, $\theta=\partial_\mu u^\mu$ is the expansion scalar, $\nabla^\mu=\Delta^{\mu\nu}\partial_\nu$ denotes the spatial gradient, and $\sigma^{\mu\nu} = \Delta^{\mu\nu}_{\alpha\beta} \nabla^\alpha u^\beta$ is the shear tensor. Since $\mathcal{E}$, $\mathcal{P}$ and $n_f$ are functions of  $\alpha \equiv \frac{\mu}{T}$ and $\beta \equiv \frac{1}{T}$, which vary with space and time, their derivatives can be expressed using the chain rule. In particular, the co-moving derivatives of the number density, energy density, and pressure are given by
\begin{align}
    \dot{n_f} &= \frac{\partial n_f}{\partial \alpha} \dot{\alpha} + \frac{\partial n_f}{\partial \beta} \dot{\beta}, \label{n_f_dot}
    \\
    \dot{\mathcal{E}} &= \frac{\partial \mathcal{E}}{\partial \alpha} \dot{\alpha} + \frac{\partial \mathcal{E}}{\partial \beta} \dot{\beta}, \label{E_dot}
    \\
    \nabla^\mu \mathcal{P} &=  \frac{\partial \mathcal{P}}{\partial \alpha} \nabla^\mu \alpha + \frac{\partial \mathcal{P}}{\partial \beta} \nabla^\mu \beta \label{nabla_P} = -\frac{1}{\beta} (\mathcal{E} + \mathcal{P}) \nabla^\mu \beta + \frac{n_f}{\beta} \nabla^\mu \alpha.
\end{align}
Substituting Eqs.~(\ref{n_f_dot}),~(\ref{E_dot}), and~(\ref{nabla_P}) into the hydrodynamic evolution equations, one can obtain the following expressions for the co-moving derivatives of the thermodynamic variables and fluid four-velocity:
\begin{align}
    \dot{\alpha} &= \frac{1}{\mathcal{D}} \left[-\frac{\partial \mathcal{E}}{\partial \beta}(n_f\theta + \partial_\mu V^\mu) + \frac{\partial n_f}{\partial \beta}\{ (\mathcal{E} + \mathcal{P} + \Pi)\theta - \pi^{\mu\nu}\sigma_{\mu\nu} \}\right], \label{alpha_dot}
    \\
    \dot{\beta} &= \frac{1}{\mathcal{D}} \left[\frac{\partial \mathcal{E}}{\partial \alpha}(n_f\theta + \partial_\mu V^\mu) - \frac{\partial n_f}{\partial \alpha}\{ (\mathcal{E} + \mathcal{P} + \Pi)\theta - \pi^{\mu\nu}\sigma_{\mu\nu} \}\right], \label{beta_dot}
    \\
    \dot{u}^\alpha &= \frac{1}{(\mathcal{E} + \mathcal{P} + \Pi)\beta} \Bigr[n_f \nabla^\alpha \alpha - (\mathcal{E} + \mathcal{P}) \nabla^\alpha \beta + \beta \nabla^\alpha \Pi - \beta \Delta^\alpha_\nu \partial_\mu \pi^{\mu\nu} - \beta qBb^{\alpha\lambda} V_{f,\lambda} \Bigr] , \label{u_dot}
\end{align}
where,
\begin{equation*}
    \mathcal{D} = \frac{\partial n_f}{\partial \alpha} \frac{\partial \mathcal{E}}{\partial \beta} - \frac{\partial n_f}{\partial \beta} \frac{\partial \mathcal{E}}{\partial \alpha} .
\end{equation*}
To establish the connection between the hydrodynamic variables and their microscopic kinetic description, the energy-momentum tensor and particle-number four-current can be expressed in terms of the corresponding distribution functions as
\begin{align}
    \mathcal{T}_f^{\mu\nu} &= \int dP_q \ p^\mu p^\nu (f_q^+ + f_q^-) + \int dP_g \ p^\mu p^\nu f_g  \ + \mathcal{B}(\alpha, \beta) g^{\mu\nu} ,
    \\
    \mathcal{N}^\mu &= \int dP_q \ p^\mu (f^+_q - f^-_q) .
\end{align}
Here, $f^+_q$, $f^-_q$, $f_g$ are the distribution functions of quarks, anti-quarks, and gluons, respectively. The integration measure for quarks/anti-quarks is defined as $dP_q = \frac{d_q d^3 \textbf{p}}{(2\pi)^3 p^0}$, where $d_q$ is the quark/anti-quark degeneracy factor. Similarly, the gluon integration measure is $dP_g = \frac{d_g d^3 \textbf{p}}{(2\pi)^3 p^0}$, with $d_g$ as the gluon degeneracy factor. For a system away from thermal equilibrium, the distribution functions can be decomposed into equilibrium and non-equilibrium contributions as,
\begin{align}
    f^\pm_q = f_{q\,\,(0)}^\pm + \delta f^\pm_q, && f_g = f_{g\,\,(0)} + \delta f_g, 
\end{align}
where $f_{q\,\,(0)}^\pm$, $f_{g\,\,(0)}$ represent the equilibrium distribution functions of quarks, anti-quarks, and gluons, respectively. These equilibrium distributions take the form,
\begin{align}
    f_{q\,\,(0)}^\pm &= \exp(-\beta (u \cdot p) \pm \alpha), \qquad \text{for quarks and anti quarks}, \\
    f_{g\,\,(0)} &= \exp(-\beta (u \cdot p)),\,\,\,\,\,\,\,\,\,\,\,\qquad  \text{for gluons}.
\end{align}
The quantities $\delta f^\pm_q$, $\delta f_g$ describe the deviation from equilibrium of quarks/anti-quarks and gluons, respectively. The equilibrium thermodynamic quantities can be obtained by evaluating the energy-momentum tensor and particle-number four-current using the equilibrium distribution functions. Accordingly, the energy density, thermal pressure, and net particle-number density are given by,
\begin{align}
    \mathcal{E} &= u_\mu u_\nu  \left( \int dP_q \ p^\mu p^\nu (f_{q\,\,(0)}^+ + f_{q\,\,(0)}^-) +  \int dP_g \ p^\mu p^\nu f_{g\,\,(0)} \right) + \mathcal{B}(\alpha, \beta), \label{E_kin}
    \\
    \mathcal{P} &= -\frac{\Delta_{\mu\nu}}{3} \left( \int dP_p \ p^\mu p^\nu (f_{q\,\,(0)^+} + f_{q\,\,(0)}^-) + \int dP_g \ p^\mu p^\nu f_{g\,\,(0)} \right) - \mathcal{B}(\alpha, \beta), 
    \\
    n_f &= u_\mu \int dP_q \ p^\mu (f_{q\,\,(0)}^+ - f_{q\,\,(0)}^-). \label{n_f_kin}
\end{align}
Upon evaluating these momentum integrals, these quantities can be expressed in terms of the thermodynamic integrals  $I^{(r)\pm}_{nq}$ as follows,
\begin{align}
    n_f &= \ I^{(0)-}_{10} (m),
    \\
    \mathcal{E} &= I^{(0)+}_{20} (m) + I^{(0)}_{20} (M) + \mathcal{B},
    \\
    \mathcal{P} &= - I^{(0)+}_{21} (m) - I^{(0)}_{21} (M) - \mathcal{B},
\end{align}
with $m(\mu,T)$ and $M(\mu,T)$ as effective masses of the quasi-quark/anti-quark and quasi-gluon. The thermodynamic integral $I^{(r)\pm}_{nq}$ can be expressed in terms of modified Bessel function $K_n(z)$ as described in Appendix~\ref{Relations}. Consequently, the thermodynamic quantities take the explicit forms,
\begin{align}
    n_f &= \frac{d_q T^3 z^2_q}{\pi^2} \sinh (\alpha) K_2(z_q),\label{thermo1}
    \\
    \mathcal{E} &= \frac{d_q T^4 z^2_q}{\pi^2} \cosh (\alpha) \left[3 K_2(z_q) + z_q K_1(z_q)\right] + \frac{d_g T^4 z^2_g}{2 \pi^2} \left[3 K_2(z_g) + z_g K_1(z_g)\right] +  \mathcal{B}(\alpha, \beta), \label{E_kin2}
    \\
    \mathcal{P} &= \frac{d_q T^4 z^2_q}{\pi^2} \cosh (\alpha) K_2(z_q) + \frac{d_g T^4 z^2_g}{2 \pi^2} K_2(z_g) - \mathcal{B}(\alpha, \beta). \label{P_kin2}
\end{align}
Here, $z_q=m(\mu,T)/T$ and $z_g=M(\mu,T)/T$ are the dimensionless effective masses of quarks and gluons, respectively. The dissipative quantities, namely the shear-stress tensor, bulk pressure, and number diffusion current can be expressed in form of the non-equilibrium part of distribution function as,
\begin{align}
    V^\mu &= \Delta^\mu_\nu \int dP_q \ p^\nu (\delta f^+_q - \delta f^-_q) \label{Vmu},
    \\
    \pi^{\mu \nu} &= \Delta^{\mu \nu}_{\alpha \beta} \left( \int dP_q \ p^\alpha p^\beta (\delta f^+_q + \delta f^-_q) + \int dP_g \ p^\alpha p^\beta \delta f_g \right) \label{pi},\\
    \Pi &= - \frac{\Delta_{\mu\nu}}{3} \left(\int dP_q \ p^\mu p^\nu (\delta f^+_q + \delta f^-_q) + \int dP_g \ p^\mu p^\nu \delta f_g \right) \label{Pi}.
\end{align}
The non-equilibrium corrections to the distribution functions, $\delta f^\pm_q$ and $\delta f_g$, are determined by solving the relativistic Boltzmann equation. In the following, we employ the kinetic theory framework to obtain the dissipative transport coefficients governing the evolution of the QCD medium in the presence of an external magnetic field.
\subsection{Effective Covariant Boltzmann Equation for Magnetized QCD Medium} \label{Boltz_Eqn}
The relativistic Boltzmann equation in the presence of a non-zero four-force $\mathcal{F}^\mu$ takes the general form as,
\begin{align}
    p^\mu \partial_\mu f + \mathcal{F}^\mu \frac{\partial f}{\partial p^\mu} = C[f] ,
\end{align}
where ${f}$ denotes the phase-space distribution function, with $f = f^\pm_q$ for quarks/anti-quarks and $f = f_g$ for gluons, while $C[{f}]$ is the collision kernel. The first term on the left-hand side describes the free streaming of particles in phase space, whereas the second accounts for the effect of external and medium-induced forces on their motion. In a QCD medium, particles interact with one another and may also couple to an external magnetic field. 
While quarks and anti-quarks carry electric charge and therefore interact with the magnetic field, gluons are electrically neutral and do not couple directly to it. Accordingly, the force acting on quarks and anti-quarks receives contributions from both the medium-induced interaction (mean-field force) $\mathcal{F}^\mu_\text{int} = m(\partial^\mu m)$ and the magnetic field (Lorentz force\footnote{The Lorentz force on anti-quarks has the opposite sign}) $\mathcal{F}^\mu_B = -qBb^{\mu\nu}p_\nu$. For gluon $\mathcal{F}^\mu_\text{int} = M(\partial^\mu M)$. It is important to emphasize that the medium-induced interaction force represents the mean-field contribution arising from the space-time dependence of the effective quasiparticle masses. 
The collision term generally involves nonlinear interactions among the particles, making the Boltzmann equation difficult to solve. In the present analysis, we employ the relaxation-time approximation (RTA) proposed by Anderson and Witting for relativistic systems~\cite{Anderson1974466}. In this approximation, the collision term is written as, $C[{f}] = -\frac{u \cdot p}{\tau_c} \Delta {f}$, where $\Delta {f} = {f} - {f}_0$ denotes the deviation from the equilibrium distribution function and $\tau_c$ is the thermal relaxation time. By incorporating both QCD medium interaction and magnetic field contributions to the force, the RTA Boltzmann equations for quarks and anti-quarks become:
\begin{equation}
    p^\mu \partial_\mu f^\pm_q + m (\partial^\mu m) \frac{\partial f^\pm_q}{\partial p^\mu} \mp q B b^{\mu\nu} p_\nu \frac{\partial f^\pm_q}{\partial p^\mu} = -\frac{u \cdot p}{\tau_c} (f^\pm_q - f^\pm_{q\,\,(0)}) . \label{BE_qrk}
\end{equation}
Similarly, for gluons, which do not couple directly to the magnetic field, we obtain
\begin{equation}
    p^\mu \partial_\mu f_g + M (\partial^\mu M) \frac{\partial f_g}{\partial p^\mu} = -\frac{u \cdot p}{\tau_c} (f_g - f_{g\,\,(0)}). \label{BE_gln}
\end{equation}
The non-equilibrium distribution function can be obtained by using the iterative Chapman-Enskog solution of the RTA Boltzmann equation. Accordingly, we write,
\begin{align}
    \delta f^\pm_q = f^\pm_{q\,\,(1)} + f^\pm_{q\,\,(2)} + ..... \ , && \delta f_g = f_{g\,\,(1)} + f_{g\,\,(2)} + ..... \ , 
\end{align}
where $f^\pm_n$ and $f_n$ denote the $n$th-order corrections to the equilibrium distribution functions for quarks/anti-quarks and gluons, respectively. Substituting these expansions into Eqs.~(\ref{BE_qrk}) and~(\ref{BE_gln}), we obtain the following iterative relations for higher-order corrections to the distribution function for quarks/anti-quarks and gluons as follows,
\begin{align}
    f^\pm_{q\,\,(n+1)} &= -\frac{\tau_c}{(u \cdot p)} \left[p^\mu \partial_\mu + m(\partial^\mu m) \frac{\partial}{\partial p^\mu} \mp q B b^{\mu\nu} p_\nu \frac{\partial}{\partial p^\mu} \right]f^\pm_{q\,\,(n)} , \label{f_pm_q}
    \\
    f_{g\,\,(n+1)} &= -\frac{\tau_c}{(u \cdot p)} \left[p^\mu \partial_\mu + M(\partial^\mu M) \frac{\partial}{\partial p^\mu} \right]f_{g\,\,(n)}, \label{f_g}
\end{align}
where $n = 0,1,2...$. Different values of $n$ will give the $ n$th-order distribution function, for example, $f^\pm_{q\,\,(1)}$ and $f^\pm_{q\,\,(2)}$ denote the first-order and second-order corrections to the distribution function for quarks and anti-quarks, respectively.
\subsection{Evolution Equation of Diffusion current}
The first-order correction to the distribution function for quarks and gluons can be deduced from Eqs.~(\ref{f_pm_q}) and~(\ref{f_g}), respectively, by setting $n=1$ as,
\begin{align}
    f^\pm_{q\,\,(1)} &= -\frac{\tau_c}{u \cdot p} \left[p^\mu \partial_\mu f^\pm_{q\,\,(0)} + m(\partial^\mu m) \frac{\partial f^\pm_{q\,\,(0)}}{\partial p^\mu} \right] ,\label{f_q_1}
    \\
    f_{g\,\,(1)} &= -\frac{\tau_c}{(u \cdot p)} \left[p^\mu \partial_\mu f_{g\,\,(0)} + M(\partial^\mu M) \frac{\partial f_{g\,\,(0)}}{\partial p^\mu} \right] . \label{f_g_1}
\end{align}
Note that the first-order corrections to the distribution functions do not explicitly depend on the magnetic field. This is consistent with the findings of Ref.~\cite{Panda:2020zhr}. The first-order diffusion current is defined as,
\begin{align}
    V^\mu_1 = \Delta^\mu_\nu \int dP_q \ p^\nu (f^+_{q\,\,(1)} - f^-_{q\,\,(1)}). \label{Vf}
\end{align}
Substituting Eq.~(\ref{f_q_1}) into Eq.~(\ref{Vf}), we obtain
\begin{align}\label{first1}
    &V^\mu_1 = \tau_c \beta_V \nabla^\mu \alpha, && \text {with}\,\,\,\,\,\,\, \beta_V = \frac{n_f}{\mathcal{E} + \mathcal{P}} I^{(0)-}_{21} - I^{(1)+}_{21}.
\end{align}
The thermal medium effects are incorporated through the thermodynamic quantities as well as the thermodynamic integrals. Similarly, substituting the first-order corrections to the quark and gluon distribution functions into Eqs.~(\ref{pi}) and~(\ref{Pi}), respectively, we obtain the first-order shear-stress tensor, and bulk pressure, given by
\begin{align}\label{first2}
    \pi^{\mu\nu}_1 &= 2 \tau_c \beta_\pi \sigma^{\mu\nu}, \\
    \Pi_1 &= -\tau_c \beta_\Pi \theta,\label{first3}
\end{align}
The coefficients $\beta_\pi$ and $\beta_\Pi$ take the forms as,
\begin{align}
    &\beta_\pi = \beta \left( I^{(1)+}_{42} + I^{(1)}_{42} \right),\\
    &\beta_\Pi = \left(I^{(0)+}_{31} + I^{(0)}_{31} \right) \mathcal{X} - I^{(0)-}_{21}\mathcal{Y} + \frac{5}{3} \beta  \left(I^{(1)+}_{42} + I^{(1)}_{42} \right) 
    + I^{(1)+}_{21} m \left(\frac{\partial m}{\partial \beta} \mathcal{X} + \frac{\partial m}{\partial \alpha} \mathcal{Y} \right) + I^{(1)}_{21} M \left(\frac{\partial M}{\partial \beta} \mathcal{X} + \frac{\partial M}{\partial \alpha} \mathcal{Y} \right),
\end{align}
respectively. The quantities $\mathcal{X}$ and $\mathcal{Y}$ can be written as,
\begin{align} 
    \mathcal{X} = \frac{1}{\mathcal{D}} \left[\frac{\partial \mathcal{E}}{\partial \alpha}n_f - \frac{\partial n_f}{\partial \alpha} (\mathcal{E} + \mathcal{P}) \right],  &&
    \mathcal{Y} = \frac{1}{\mathcal{D}} \left[\frac{\partial n_f}{\partial \beta}(\mathcal{E} + \mathcal{P})-\frac{\partial \mathcal{E}}{\partial \beta}n_f\right].\label{X_Y}
\end{align}
Similar to the first-order expansion, the second-order correction to the distribution function for quarks/anti-quarks can be written as,
\begin{multline}
    f^\pm_{q\,\,(2)} = \frac{\tau_c}{u \cdot p} p^\gamma \partial_\gamma \left[\frac{\tau_c}{u \cdot p} p^\lambda \partial_\lambda f^\pm_{q\,\,(0)} \right] 
    + \frac{\tau_c}{u \cdot p} p^\gamma \partial_\gamma \left[\frac{\tau_c}{u \cdot p} m (\partial^\lambda m) \frac{\partial f^\pm_{q\,\,(0)}}{\partial p^\lambda} \right] 
    + \frac{\tau_c}{u \cdot p} m (\partial^\gamma m) \frac{\partial}{\partial p^\gamma} \left[\frac{\tau_c}{u \cdot p} p^\lambda \partial_\lambda f^\pm_{q\,\,(0)} \right]
    \\
    + \frac{\tau_c}{u \cdot p} m (\partial^\gamma m) \frac{\partial}{\partial p^\gamma} \left[ \frac{\tau_c}{u \cdot p} m (\partial^\lambda m) \frac{\partial f^\pm_{q\,\,(0)}}{\partial p^\lambda} \right]
    \mp \frac{\tau_c}{u \cdot p} q B b^{\sigma\gamma} p_\gamma \frac{\partial}{\partial p^\sigma} \left[\frac{\tau_c}{u \cdot p} p^\lambda \partial_\lambda f^\pm_{q\,\,(0)} \right]
    \mp\frac{\tau_c}{u \cdot p} q B b^{\sigma\gamma} p_\gamma \frac{\partial}{\partial p^\sigma} \left[\frac{\tau_c}{u \cdot p} m (\partial^\lambda m) \frac{\partial f^\pm_{q\,\,(0)}}{\partial p^\lambda} \right] \label{f_pm_2}.
\end{multline}
The first term is the standard RTA-based second-order correction to the distribution function. Second, third, and fourth terms have explicit space-time dependence of the derivative of the effective mass, which arises solely due to the QCD medium interactions. The fifth term arises explicitly due to the magnetic field in the medium, and the last term couples the magnetic field effect with thermal QCD medium effects. The expression for the second-order diffusion current can be expressed as,
\begin{equation}
    V^\mu_2 = \Delta^\mu_\nu \int dP_q \ p^\nu (f^+_{q\,\,(2)} - f^-_{q\,\,(2)}) . \label{V_mu_2}
\end{equation}
Substituting Eq.~(\ref{f_pm_2}) in Eq.~(\ref{V_mu_2}), we obtain 
\begin{equation}
\begin{split}
    \dot{V}^{\langle \mu \rangle} &= - \frac{V^{\mu}}{\tau_c} - \tau_c \Delta^\mu_\alpha D \left[\frac{n_f q B b^{\alpha \lambda} V_\lambda}{\mathcal{E} + \mathcal{P}} \right] - \tau_c \left( I^{(0)-}_{20} \mathcal{X} - I^{(1)+}_{20} \mathcal{Y} + \beta I^{(0)-}_{21} \right) \dot{u}^\mu \theta - \left[\frac{n_f}{\mathcal{E} + \mathcal{P}} I^{(0)-}_{21} - I^{(1)+}_{21}\right] \nabla^\mu \alpha
    \\
    &-\tau_c \nabla^\mu \left[\left(I^{(0)-}_{21} \mathcal{X} - I^{(1)+}_{21} \mathcal{Y} + \frac{5}{3} \beta I^{(2)-}_{42}\right) \theta\right] + 2\tau_c \Delta^\mu_\alpha \nabla_\gamma \Bigl[\beta I^{(2)-}_{42} \sigma^{\gamma \alpha} \Bigr] - \tau_c \left[\frac{n_f}{\mathcal{E} + \mathcal{P}} I^{(0)-}_{21} -I^{(1)+}_{21} \right] (\nabla^\gamma \alpha) \omega^\mu_\gamma 
    \\
    &-\tau_c \left[\frac{4}{3} \left(\frac{n_f}{\mathcal{E} + \mathcal{P}} I^{(0)-}_{21} -I^{(1)+}_{21} \right) + \frac{5}{3}\left(\frac{n_f}{\mathcal{E} + \mathcal{P}} I^{(2)-}_{42} -I^{(3)+}_{42} \right) \right] (\nabla^\mu \alpha) \theta + \tau_c \left[\frac{4}{3} I^{(0)-}_{21} + \frac{5}{3} I^{(2)-}_{42}\right] \theta \left[\frac{\beta q B b^{\mu\lambda} V_\lambda}{\mathcal{E} + \mathcal{P}} \right]
    \\
    &-\tau_c \left[\left(\frac{n_f}{\mathcal{E} + \mathcal{P}} I^{(0)-}_{21} -I^{(1)+}_{21} \right) + 2 \left(\frac{n_f}{\mathcal{E} + \mathcal{P}} I^{(2)-}_{42} -I^{(3)+}_{42} \right) \right] (\nabla^\gamma \alpha) \sigma^\mu_\gamma + \tau_c \left(I^{(0)-}_{21} + 2 I^{(2)-}_{42} \right) \sigma^\mu_\gamma \left[\frac{\beta q B b^{\gamma\lambda} V_\lambda}{\mathcal{E} + \mathcal{P}} \right]
    \\
    &+ \tau_c I^{(0)-}_{21} \omega^\mu_\gamma \left[\frac{\beta q B b^{\gamma\lambda} V_\lambda}{\mathcal{E} + \mathcal{P}} \right] + \tau_c \left[\frac{n_f}{\mathcal{E} + \mathcal{P}} I^{(1)+}_{21} -I^{(2)-}_{21} \right] q B b^{\gamma\mu} (\nabla_\gamma \alpha) + \tau_c I^{(1)+}_{21} \beta q B b^{\gamma\mu}\left[\frac{\nabla_\gamma \Pi}{\mathcal{E} + \mathcal{P}}-\frac{\Pi \dot{u}_\gamma}{\mathcal{E} + \mathcal{P}} -\frac{\Delta^\sigma_\gamma \partial^\kappa \pi_{\kappa\sigma}}{\mathcal{E} + \mathcal{P}} \right]
    \\
    &+ \tau_c I^{(2)-}_{21} \beta m \left( \frac{\partial m}{\partial \beta} \mathcal{X} + \frac{\partial m}{\partial \alpha} \mathcal{Y} \right) \dot{u}^\mu \theta - \tau_c I^{(2)-}_{21} \nabla^\mu \left[\beta m \left(\frac{\partial m}{\partial \beta} \mathcal{X} + \frac{\partial m}{\partial \alpha} \mathcal{Y} \right) \theta \right] - 2 \tau_c I^{(1)-}_{21} \beta m \left( \frac{\partial m}{\partial \beta} \mathcal{X} + \frac{\partial m}{\partial \alpha} \mathcal{Y} \right) \theta \frac{\beta q B b^{\mu \lambda} V_\lambda}{\mathcal{E} + \mathcal{P}}
    \\
    &+ \tau_c \beta m \left( \frac{\partial m}{\partial \beta} \mathcal{X} + \frac{\partial m}{\partial \alpha} \mathcal{Y} \right) \left[ \frac{n_f}{\mathcal{E} + \mathcal{P}}I^{(1)-}_{21} - I^{(2)+}_{21} \right] (\nabla^\mu \alpha) \theta - \tau_c I^{(1)-}_{21}\beta m \left( \frac{\partial m}{\partial \beta} \mathcal{X} + \frac{\partial m}{\partial \alpha} \mathcal{Y} \right) \theta \frac{\beta q B b^{\mu\lambda} V_\lambda}{\mathcal{E} + \mathcal{P}}
    \\
    &+ \tau_c I^{(1)-}_{21} \beta m \left( \frac{\partial m}{\partial \beta} \mathcal{X} + \frac{\partial m}{\partial \alpha} \mathcal{Y} \right) \frac{n_f}{\mathcal{E} + \mathcal{P}} (\nabla^\mu \alpha) \theta - \tau_c \Bigl(I^{(3)+}_{21} + I^{(2)+}_{21} \beta \Bigr) m \left( \frac{\partial m}{\partial \beta} \mathcal{X} + \frac{\partial m}{\partial \alpha} \mathcal{Y} \right) (\nabla^\mu \alpha)\theta
    \\
    &+ 2 \tau_c \beta^2 I^{(2)-}_{21} m \frac{\partial m}{\partial \beta} 
    \dot{u}^\lambda \sigma^\mu_\lambda - 2 \tau_c \beta I^{(2)-}_{21} m \left( \frac{n_f}{\mathcal{E} + \mathcal{P}} \frac{\partial m}{\partial \beta} + \frac{\partial m}{\partial \alpha} \right) \sigma^\mu_\lambda (\nabla^\lambda \alpha) + 2\tau_c I^{(2)-}_{21} \beta m \frac{\partial m}{\partial \beta} \sigma^\mu_\lambda \frac{\beta q B b^{\lambda \rho} V_\rho}{\mathcal{E} + \mathcal{P}}
    \\
    &+ \frac{2}{3} \tau_c \beta^2 I^{(2)-}_{21} m \frac{\partial m}{\partial \beta} \dot{u}^\mu \theta - \frac{2}{3} \tau_c \beta I^{(2)-}_{21} m \left( \frac{n_f}{\mathcal{E} + \mathcal{P}} \frac{\partial m}{\partial \beta} + \frac{\partial m}{\partial \alpha} \right) (\nabla^\mu \alpha) \theta + \frac{2}{3} \tau_c I^{(2)-}_{21} \beta m \frac{\partial m}{\partial \beta} \theta \frac{\beta q B b^{\mu \lambda} V_\lambda}{\mathcal{E} + \mathcal{P}}. 
\end{split}
\end{equation}
The total diffusion current is composed of both first and second-order terms, $i.e.$, $V^\mu = V^\mu_1 + V^\mu_2$. By rearranging the terms according to their respective tensorial structures, we obtain the second-order evolution equation for the diffusion current as follows:
\begin{equation} \label{V_mu_2nd}
\begin{split}
    \dot{V}^{\langle \mu \rangle} &= - \frac{V^\mu}{\tau_c} - V_\nu \omega^{\nu\mu} - \lambda_{VV} V^\nu \sigma^\mu_\nu - \delta_{VV} V^\mu \theta + \lambda_{V\Pi} \Pi \nabla^\mu \alpha - \lambda_{V\pi} \pi^{\mu\nu} \nabla_\nu \alpha - \tau_{V\pi} \pi^\mu_\nu \dot{u}^\nu + \tau_{V\Pi} \Pi \dot{u}^\mu + l_{V\pi} \Delta^{\mu\nu} \partial_\gamma \pi^\gamma_\nu
    \\
   & - l_{V\Pi} \nabla^\mu \Pi + \beta_V \nabla^\mu \alpha - qB \delta_{VB} b^{\mu\gamma}V_\gamma + \tau_c q B l_{V \pi B} b^{\sigma \mu} \partial^\kappa \pi_{\kappa\sigma} + \tau_c q B \tau_{V\Pi B} b^{\gamma\mu} \Pi \dot{u}_\gamma - \tau_c q B l_{V \Pi B} b^{\gamma\mu} \nabla_\gamma \Pi  
    \\
    & - \tau_c q \delta_{VVB} B b^{\mu\nu} V_\nu \theta - \tau_c q \lambda_{VVB} B b^{\gamma\nu} V_\nu \sigma^\mu_\gamma - \tau_c q \rho_{VVB} B b^{\gamma\nu} V_\nu \omega^\mu_\gamma - \tau_c q \tau_{VVB} \Delta^\mu_\gamma D (B b^{\gamma \nu} V_\nu ).
\end{split}
\end{equation}
The explicit expressions for the second-order transport coefficients, in terms of thermodynamic integrals, are presented in Appendix~\ref{Vmu_transp_coeff}. The coefficients $\tau_{V\pi}$, $\lambda_{V\pi}$, and $\lambda_{VVB}$ involve derivatives of $l_{V\pi}$, whereas $\tau_{V\Pi}$, and $\lambda_{V\Pi}$ involve derivatives of $l_{V\Pi}$. The coefficient $\delta_{VVB}$ contains contributions involving derivatives of both $\tau_{VVB}$ and $l_{V\Pi}$. Notably, all the transport coefficients are modified by the inclusion of realistic QCD EoS effects. We have verified that, in the ideal EoS limit, our results reduce to the corresponding transport coefficients obtained in Ref.~\cite{Panda:2020zhr}. This demonstrates the consistency of our results with the existing literature. Furthermore, the effective-mass parametrization plays an important role in quantifying the thermal medium effects on the transport coefficients.

    \begin{figure}
        \centering
        \hspace{20pt}
        \begin{minipage}[t]{0.3\linewidth}
            \includegraphics[width=\linewidth]{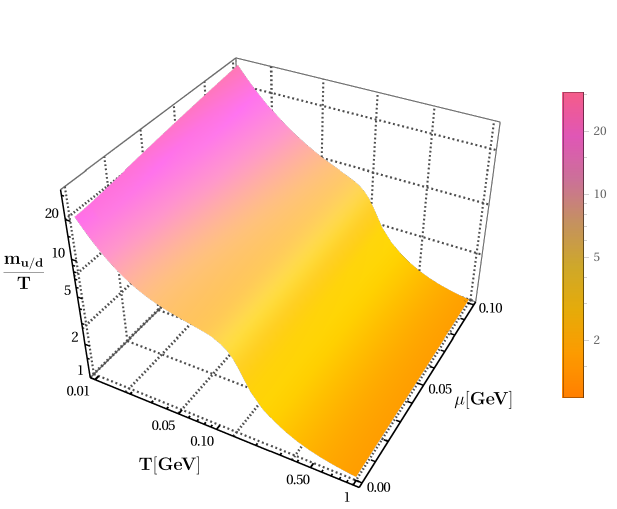}
        \end{minipage}
        \hfill
        \begin{minipage}[t]{0.3\linewidth}
            \includegraphics[width=\linewidth]{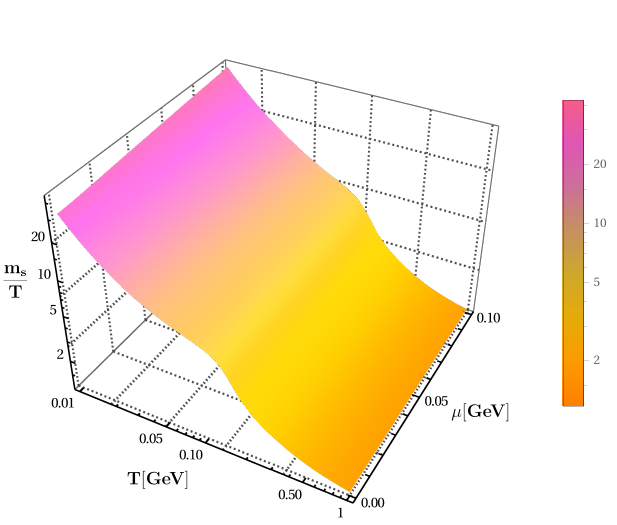}
        \end{minipage}
        \hfill
        \begin{minipage}[t]{0.3\linewidth}
            \includegraphics[width=\linewidth]{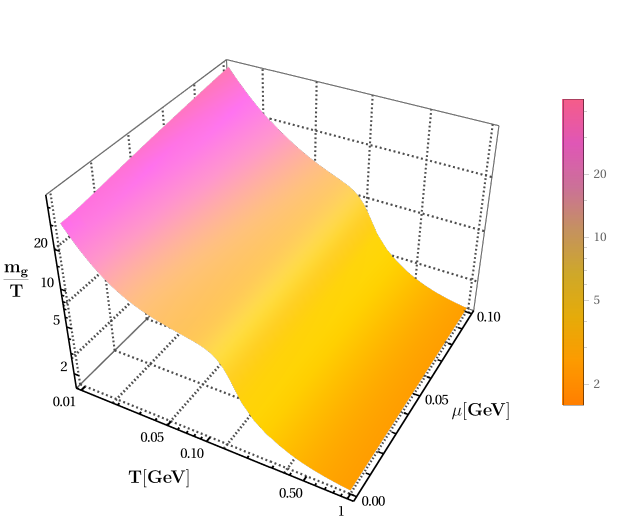}
        \end{minipage}
        \caption{Effective masses of (i) up/down quarks, (ii) strange quarks, and (iii) gluons extracted using Model 1.}
        \label{fig1}
    \end{figure}
\section{Interplay of EoS and Magnetic Field} \label{Result}
\subsection{Effective Masses of Quarks and Gluons}\label{IIIA}
In this section, we consider different prescriptions for determining the effective quasiparticle masses of quarks and gluons. For convenience, we denote the effective masses of quarks and gluons, $m(\mu,T)$ and $ M(\mu,T)$, collectively by $(m_\mathrm{eff})_a$ where $a$ denotes the particle species. We investigate how different effective-mass parametrizations influence the hydrodynamic evolution of the QCD medium in the presence of an external magnetic field. Following Ref.~\cite{Rocha:2024rce}, the temperature-dependent effective masses are determined by matching the entropy density obtained from microscopic theory to that extracted from lattice QCD. For this purpose, we employ the lattice-QCD EoS for $(2+1)$ quark flavors calculated by the Wuppertal--Budapest Collaboration~\cite{Borsanyi:2010cj}. The trace anomaly, $\mathcal{I}(T)$, is parametrized as,
\begin{equation}
    \mathcal{I}(T) = T^4 \exp \left(-\frac{h_1}{t} -\frac{h_2}{t^2} \right) \left[ \frac{h_0}{1 + h_3 t^2} + l_0 \frac{\tanh(l_1t + l_2) + 1}{1 + g_1 t + g_2 t^2}\right],
\end{equation}
where $t=T/(0.2 \text{ GeV})$, and parameters are $h_0 = 0.1396$, $h_1 = -0.1800$, $h_2 = 0.0350$, $h_3 = 0.01$, $l_0 = 2.76$, $l_1 = 6.79$, $l_2 = -5.29$, $g_1 = -0.47$, and $g_2 = 1.04$. The lattice-QCD  energy density $\mathcal{E}_\text{latt}$ pressure $\mathcal{P}_\text{latt}$ can be obtained from the trace anomaly using the thermodynamic relations~\cite{Tinti:2016bav, Rocha:2024rce},
\begin{align}
    \mathcal{P}_\text{latt}(T) &= T^4 \int^T_0 \frac{dT'}{T'} \frac{\mathcal{I}(T')}{T'^4},
    \\
    \mathcal{E}_\text{latt}(T) &= \mathcal{I}(T) + 3 \mathcal{P}_\text{latt}(T).
\end{align}
Subsequently, the entropy density can be calculated from lattice QCD using the relation 
\begin{equation}
    \mathcal{S}_\text{latt}(T) = \frac{\mathcal{E}_\text{latt}(T) + \mathcal{P}_\text{latt}(T)}{T}.
\end{equation}
From the microscopic theory, the entropy density can be expressed in terms of the effective quasiparticle masses of the different particle species as,
\begin{equation}
    \mathcal{S}(\mu, T) = \sum_a\frac{d_a T^3 z_a^3}{2 \pi^2} \cosh \left(\frac{\mu_a}{T} \right) K_3(z_a), \label{S_kin} 
\end{equation}
where $z_a = (m_\mathrm{eff})_a/T$. Here, $K_3(z_a)$ is the modified Bessel function of the second kind, and $d_a$ is the corresponding degeneracy factor. The sum runs over all particle species considered in the quasiparticle description. By matching the microscopic entropy density to the lattice-QCD result, the effective mass parameterizations can be determined. We consider three different prescriptions for the effective quasiparticle masses, which are described below:
  \begin{figure}
        \centering
        \includegraphics[width=0.44\linewidth]{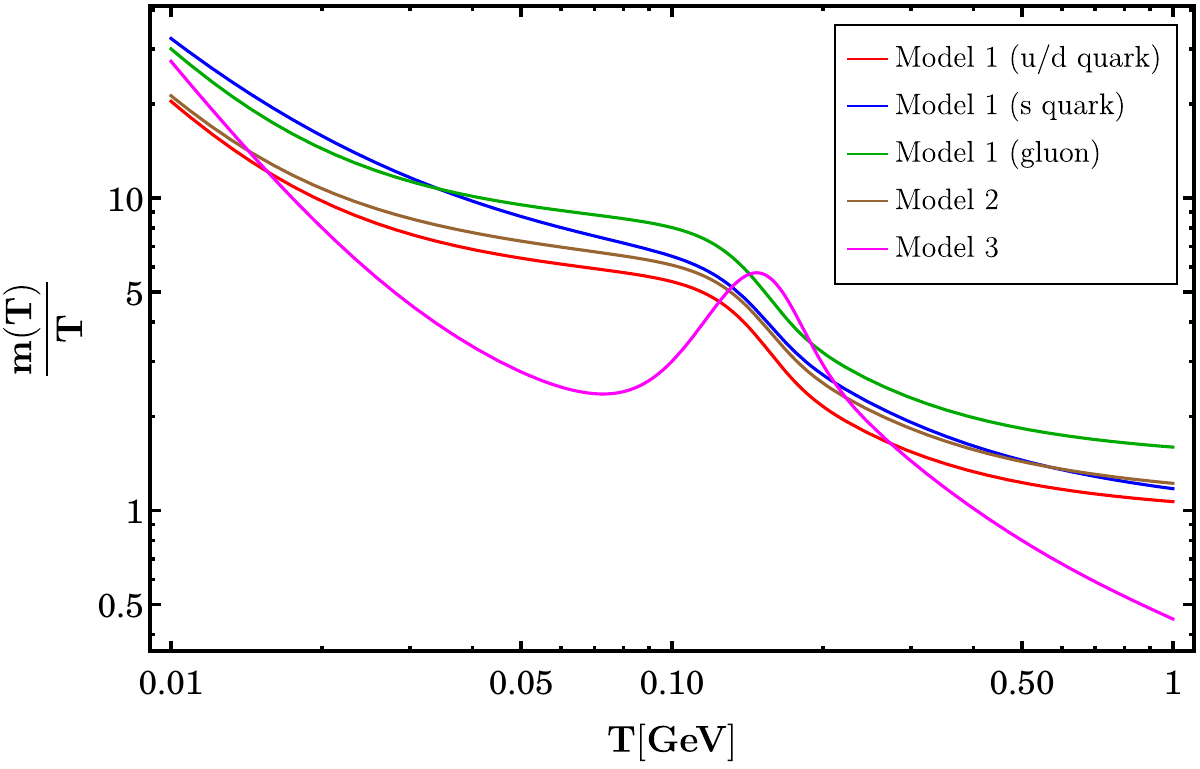}
        \caption{Temperature dependence of the effective masses obtained using the three prescriptions at $\mu=0$. }
        \label{fig2}
    \end{figure}
\begin{itemize}
    \item \textbf{Model 1: Species-dependent effective masses:} In the first model, the effective masses of up, down, and strange quarks, as well as gluons, are determined separately. Following Ref.~\cite{Bluhm:2007nu}, the effective mass squared is expressed as the sum of the square of the bare mass $m_a$  and an interaction term $\Lambda_a$, as
    \begin{equation}
        (m_\mathrm{eff})_a^2 = m^2_a + \Lambda_a \label{meff}
    \end{equation}
    Here, $a=\{u,d,s,g\}$ correspond to up, down, strange quark, and gluon, respectively. The bare mass of strange quark to be $m_{s} = 0.015 \text{ GeV}$, while up and down quark masses are set to $m_u = m_d = m_s/28.15 \text{ GeV}$. The degeneracy factors for quarks and gluons are given by $d_{u, d, s} = 2 N_c$ (for each flavor) and $d_g = 2 (N_c^2 - 1)$, with $N_c = 3$. The interaction terms for quark  and gluon are given by,  
    \begin{align}
        \Lambda_a &= 2 \left[m_a \sqrt{\frac{g_\mathrm{eff}^2}{6} \left(T^2 +\frac{\mu^2}{\pi^2} \right)} +\frac{g_\mathrm{eff}^2}{6} \left(T^2 +\frac{\mu^2}{\pi^2} \right) \right], \label{int_q} \quad a=\{u, d, s\}
        \\
        \Lambda_g &= \frac{g_{\mathrm{eff}}^2}{6} \left[ \left(N_c + \frac{N_f}{2} \right) T^2 + \frac{9 \mu^2}{2 \pi^2} \right]. \label{int_g}
    \end{align}
The effective coupling, $g_\mathrm{eff}$, is determined by matching the total entropy density obtained from the microscopic theory expressions for the different particle species [Eq.~(\ref{S_kin})] to the corresponding lattice-QCD result. Once \(g_\mathrm{eff}\) is determined, the effective masses of the up, down, and strange quarks and gluons are calculated using Eqs.~(\ref{meff})--(\ref{int_g}). The temperature and chemical-potential dependence of the effective masses for the individual quark flavors and gluons, obtained using this prescription, is illustrated in Fig.~\ref{fig1}. Although the lattice-QCD equation of state employed here is evaluated at zero chemical potential, we retain a finite chemical potential in the quasiparticle description and treat it as an independent parameter. Thus, the effective masses are calibrated using the zero-chemical-potential lattice-QCD equation of state, while their chemical-potential dependence is incorporated through the quasiparticle parametrization.
    \item \textbf{Model 2: Single-species quasiparticle effective mass:} In the second model, following Refs.~\cite{Tinti:2016bav, Rocha:2024rce}, an effective single-species parametrization is considered. This effective mass is determined numerically by matching the quasiparticle entropy density to the lattice-QCD result. The matching condition is given by
    \begin{equation}
        \frac{d_\mathrm{eff} T^3}{2 \pi^2} \left(\frac{m_\mathrm{eff}}{T} \right)^3 \cosh \left(\frac{\mu}{T} \right) K_3\left(\frac{m_\mathrm{eff}}{T} \right) = \mathcal{S}_\mathrm{latt}(T),
    \end{equation}
    where the effective degeneracy factor $d_\mathrm{eff}$ is defined as,
    \begin{equation}
        d_\mathrm{eff} = \frac{\pi^4}{180} \left(4(N_c^2 - 1) + 7 N_f N_c\right).
    \end{equation}
    \item \textbf{Model 3: Effective mass from lattice-QCD thermodynamics:} As a reference for comparison with the previous two prescriptions, we also consider a case where the mass is extracted by matching the squared speed of sound of the kinetic-theory description to that obtained from the lattice-QCD equation of state. Following Ref.~\cite{Tinti:2016bav}, the temperature-dependent mass is determined from the relation,
    \begin{equation}
        m_\mathrm{eff}(T) = T \sqrt{\frac{\mathcal{E}_\mathrm{latt} + \mathcal{P}_\mathrm{latt}}{c_s^2 \mathcal{P}_\mathrm{latt}} \left(\frac{1}{3} - c_s^2 \right)},
    \end{equation}
    where $c_s^2 = \frac{\partial \mathcal{P}_\text{latt}}{\partial \mathcal{E}_\text{latt}}$ which is numerically calculated using lattice-QCD result. This prescription provides an additional reference for examining how the choice of mass parametrization influences the transport coefficients and the resulting hydrodynamic evolution in the presence of an external magnetic field.
\end{itemize}
Figure~\ref{fig2} depicts the temperature dependence of the effective masses obtained from the three prescriptions at zero chemical potential, highlighting the differences arising from the respective mass parametrizations. Next, we analyze the sensitivity of the magnetic-field-modified transport coefficients to the effective mass parametrization.
\subsection{Magnetic field-modified transport coefficients}
%
\begin{figure}
    \centering
    \begin{minipage}[t]{0.329\linewidth}
        \includegraphics[width=\linewidth]{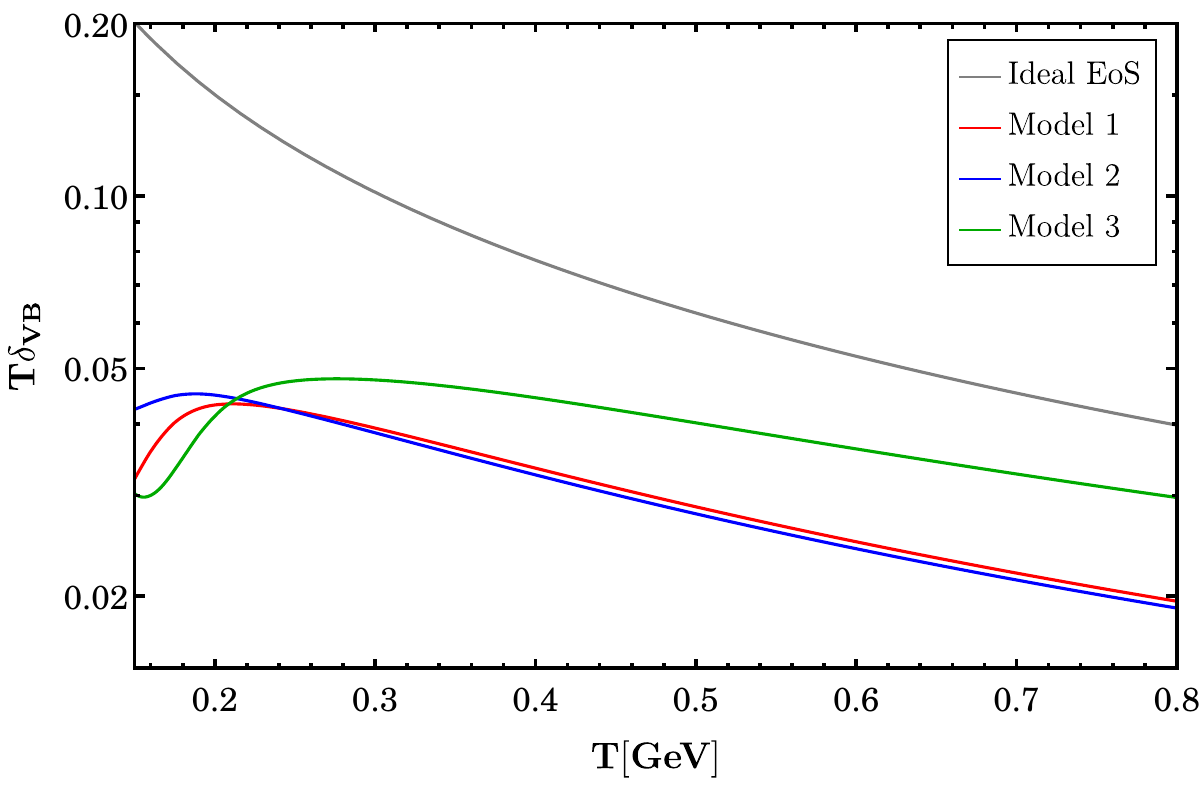}
    \end{minipage}
    \hfill
    \begin{minipage}[t]{0.329\linewidth}
        \includegraphics[width=\linewidth]{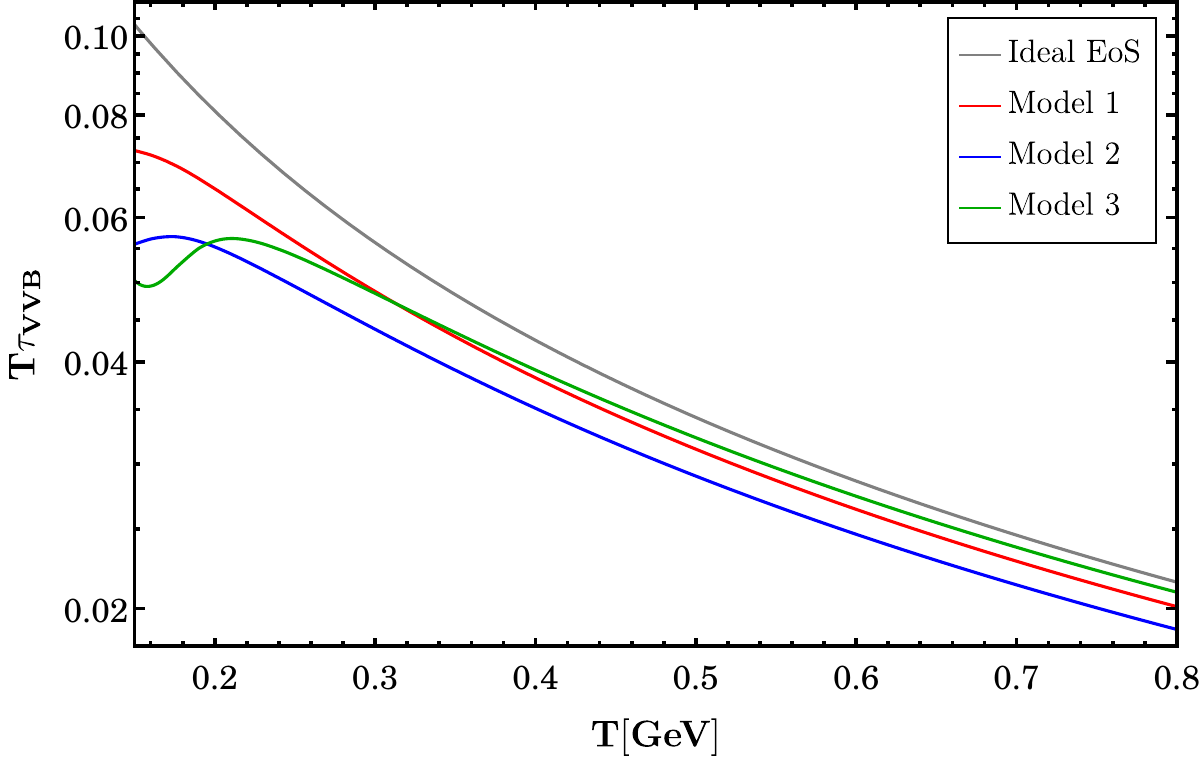}
    \end{minipage}
    \hfill
    \begin{minipage}[t]{0.329\linewidth}
        \includegraphics[width=\linewidth]{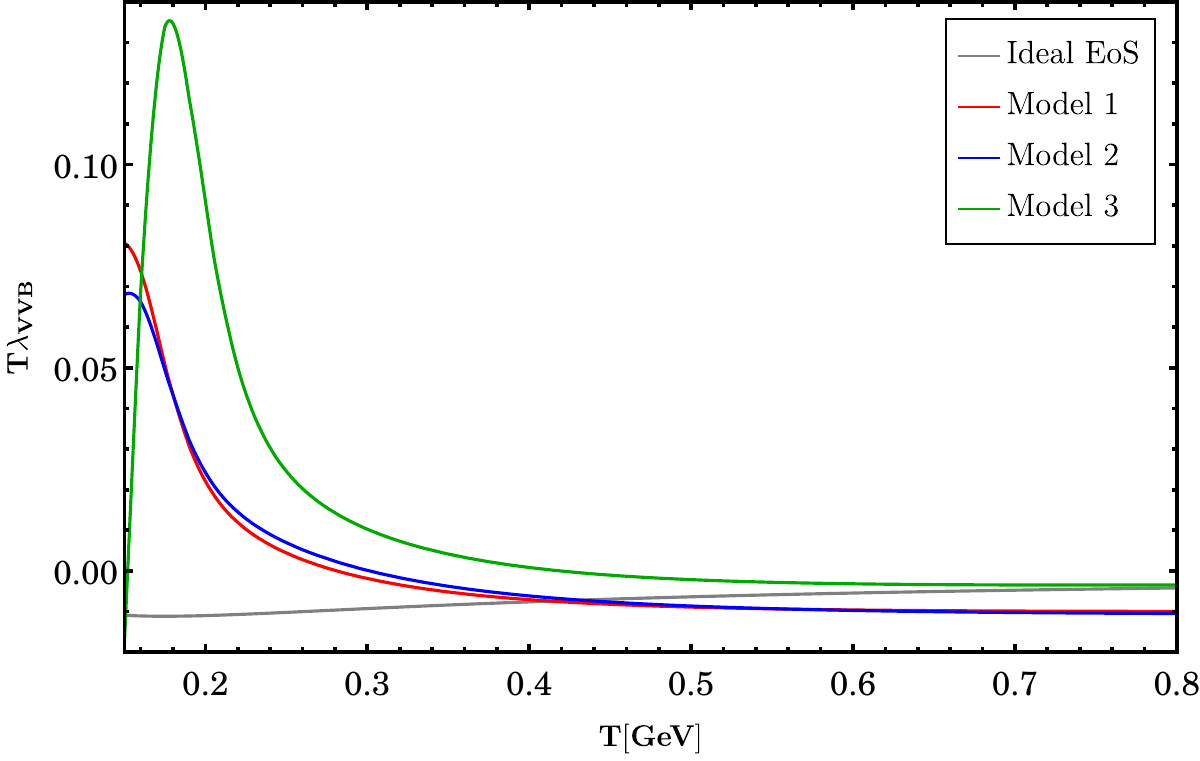}
    \end{minipage}
    \caption{Temperature behaviour of coefficients $\delta_{VB}$, $\tau_{VVB}$, and $\lambda_{VVB}$.}
    \label{1_2_order_coeff}
\end{figure}
Now, we examine how the choice of effective-mass parametrization influences the particle diffusion in the medium in the presence of a magnetic field. Although the different models are constrained by the same lattice QCD thermodynamic information, they differ in their prescription for the temperature dependence of the effective masses. Even when different effective-mass models reproduce similar equilibrium thermodynamic behavior, they can lead to different transport coefficients because of their distinct temperature and chemical-potential dependence. This provides a direct way to assess the sensitivity of the diffusion dynamics to the underlying description of the medium. The effect of thermal medium effects on the particle diffusion has recently analyzed in Ref.~\cite{Daher:2024vxk}. Here, we focus on the contributions induced by the magnetic field, particularly the sensitivity of the first-order and second-order magnetic field-modified transport coefficients associated with the diffusion transport on the thermal QCD medium effects. The form of the magnetic field-dependent transport coefficients associated with the particle number evolution as described in Eq.~(\ref{V_mu_2nd}) is obtained as follows:
\begin{align*} 
    &\delta_{VB}=\frac{1}{\beta_V} \left[\frac{n_f}{\mathcal{E} + \mathcal{P}} I^{(1)+}_{21} - I^{(2)-}_{21} \right], 
     \\
    &l_{V \pi B}= - \frac{\beta I^{(1)+}_{21}}{\mathcal{E} + \mathcal{P}},\\
    &\tau_{V \Pi B}=
        - \frac{\beta I^{(1)+}_{21}}{\mathcal{E} + \mathcal{P}},
    \\
    &l_{V \Pi B}=
        - \frac{\beta I^{(1)+}_{21}}{\mathcal{E} + \mathcal{P}},\\
         &\rho_{VVB}=- \frac{\beta I^{(0)-}_{21}}{\mathcal{E} + \mathcal{P}},\\
    &\tau_{VVB}= \frac{n_f}{\mathcal{E} + \mathcal{P}},
    \\
    &\delta_{VVB}=
         \left[\frac{\partial \tau_{VVB}}{\partial \beta} \mathcal{X} +\frac{\partial \tau_{VVB}}{\partial \alpha} \mathcal{Y} \right] - \frac{\beta}{\mathcal{E} + \mathcal{P}}\left[\frac{4}{3} I^{(0)-}_{21} + \frac{5}{3} I^{(2)-}_{42} \right] - \frac{\beta \beta_\Pi}{\mathcal{E} + \mathcal{P}} \left\{\frac{\partial}{\partial \beta} \left[\frac{1}{\beta_\Pi} \left(I^{(0)-}_{21} \mathcal{X} - I^{(1)+}_{21} \mathcal{Y} + \frac{5}{3} \beta I^{(2)-}_{42} \right)\right] \right.
        \\
        & \left. \,\,\,\,\,\,\,\,\,\,\,\,\,\,\,\,\,\,\,\,+ I^{(2)-}_{21} \frac{\partial}{\partial \beta} \left[\frac{\beta}{\beta_\Pi} m \left(\frac{\partial m}{\partial \beta} \mathcal{X} + \frac{\partial m}{\partial \beta} \mathcal{Y}\right) \right] \right\} - \frac{2}{3} I^{(2)-}_{21} \frac{\beta^2}{\mathcal{E} + \mathcal{P}} m \frac{\partial m}{\partial \beta} + 2 I^{(1)-}_{21} \frac{\beta^2}{\mathcal{E} + \mathcal{P}} m \left(\frac{\partial m}{\partial \beta} \mathcal{X} + \frac{\partial m}{\partial \alpha} \mathcal{Y} \right),\\
    &\lambda_{VVB}=\frac{2 \beta \beta_\pi}{\mathcal{E} + \mathcal{P}} \frac{\partial l_{V\pi}}{\partial \beta} - \frac{\beta}{\mathcal{E} + \mathcal{P}} \left[I^{(0)-}_{21} + 2 I^{(2)-}_{42} \right] - 2 I^{(2)-}_{21} \frac{\beta^2}{\mathcal{E} + \mathcal{P}} m \frac{\partial m}{\partial \beta},  
\end{align*}
 where $\mathcal{X}$ and $\mathcal{Y}$ are defined as in Eq.~(\ref{X_Y}).  The thermodynamic quantities of the interacting medium  $n_f$, $\mathcal{E}$, and $\mathcal{P}$ are described in Eqs.~(\ref{thermo1})-(~\ref{P_kin2}). The quantities $\beta_V$, $\beta_\pi$, and $\beta_\Pi$ are first order coefficients as defined in Eqs.~(\ref{first1})-(\ref{first3}), and higher order coefficients such as $l_{V\Pi}$ are defined in Appendix~\ref{Vmu_transp_coeff}. 
The medium effects are encoded in the transport coefficients through the thermodynamic integrals and the derivatives of the effective mass ($m\equiv m_{eff}$) with respect to the thermodynamic variables. In particular, the magnetic-field-dependent coefficients, $\delta_{VVB}$ and $\lambda_{VVB}$, contain explicit contributions involving $\partial m/\partial\beta$ and $\partial m/\partial\alpha$, in addition to the corresponding thermodynamic integrals. 

To illustrate the dependence of thermal medium effects dependence on magnetic field-modified transport coefficients, we analyzed the temperature dependence of $\delta_{VB}$, $\tau_{VVB}$, and  $\lambda_{VVB}$ in Fig.~\ref{1_2_order_coeff}. The coefficients obtained within the quasiparticle description are compared with those of a system with the ideal EoS. We observe that the inclusion of medium interactions have a significant impact on the transport properties of magnetized QCD matter. The result of the dimensionless quantity $T\delta_{VB}$ for a system with the ideal EoS remains substantially larger than the results obtained with the medium-interaction models over the entire temperature range considered. It also decreases monotonically with increasing temperature, indicating that the medium effects lead to a significant reduction of the magnitude of $\delta_{VB}$, especially at the low-temperature regime. Among the medium-interaction models, the three models exhibit noticeably different behavior. The differences between the three medium-interaction schemes can be understood from the way the effective mass enters the transport coefficient. For the coefficient $T\tau_{VVB}$, all three effective-mass schemes, as well as the constant-mass case, exhibit an overall decreasing behavior with increasing temperature. Although the qualitative temperature dependence is similar, the magnitude of the coefficient differs among the different schemes. Similar to the behavior of $T\delta_{VB}$, the inclusion of medium interactions leads to a suppression of $T\tau_{VVB}$ compared to the ideal EoS results. However, the temperature dependence of $T\lambda_{VVB}$ exhibits a notably different behavior. In contrast to the suppression observed for the other two coefficients, the incorporation of QCD medium effects leads to an enhancement of $T\lambda_{VVB}$, particularly in the low-temperature regime. This distinct behavior can be attributed, at least partly, to the explicit dependence of $T\lambda_{VVB}$ on the derivatives of the effective mass with respect to thermodynamic variables, in addition to its dependence on the thermodynamic momentum integrals. Such derivative contributions introduce an additional sensitivity to the temperature dependence of the quasiparticle mass, which can significantly influence the magnitude and temperature dependence of the transport coefficient.

\subsection{Limiting cases and consistency checks}
We first consider the weak magnetic-field regime, characterized by
$qB/T^2 \ll 1$, where $qB$ denotes the magnetic field scale and
$T$ is the temperature of the medium. 
In the presence of a magnetic field,
RTA involves two relevant expansion parameters: the gradient expansion parameter
$\tau_c T\,\partial$ and the magnetic-field-dependent parameter $\chi = \frac{qB\tau_c}{T}$,
where $\tau_c$ is the relaxation time. In the regime $qB/T^2 \ll 1$, the
magnetic-field-dependent parameter $\chi$ is suppressed relative
to the gradient expansion parameter. Consequently, up to second
order, we retain contributions of $\mathcal{O}(\partial)$ and
$\mathcal{O}(\partial^2)$, together with the mixed magnetic-field and
gradient corrections of $\mathcal{O}(\chi\,\partial)$, and terms of higher
order in these expansion parameters are neglected. Within this
power counting, the evolution of number diffusion as described in Eq.~(\ref{V_mu_2nd}) reduces to 
\begin{equation} \label{V_mu_3}
\begin{split}
    \dot{V}^{\langle \mu \rangle} &= - \frac{V^\mu}{\tau_c} - V_\nu \omega^{\nu\mu} - \lambda_{VV} V^\nu \sigma^\mu_\nu - \delta_{VV} V^\mu \theta + \lambda_{V\Pi} \Pi \nabla^\mu \alpha - \lambda_{V\pi} \pi^{\mu\nu} \nabla_\nu \alpha \\
   &- \tau_{V\pi} \pi^\mu_\nu \dot{u}^\nu + \tau_{V\Pi} \Pi \dot{u}^\mu + l_{V\pi} \Delta^{\mu\nu} \partial_\gamma \pi^\gamma_\nu
     - l_{V\Pi} \nabla^\mu \Pi + \beta_V \nabla^\mu \alpha - qB \delta_{VB} b^{\mu\gamma}V_\gamma.
\end{split}
\end{equation}
This is consistent with the result that obtained from using the 14-moment method~\cite{Denicol:2018rbw}. Further, we considered the Navier-Stokes limit by retaining only first-order
contributions in the gradient expansion. Accordingly, we keep
terms of $\mathcal{O}(\partial)$ and the mixed magnetic-field-dependent
contributions of $\mathcal{O}(\chi\,\partial)$, while discarding second-
and higher-order gradient corrections. Thus, from the right-hand
sides of Eq.~(\ref{V_mu_3}), only three terms would survive. The resulting equation for diffusion current within the RTA can be written as,
\begin{equation}
    \left(\frac{g^{\mu\nu}}{\tau_c} + q B \delta_{VB} b^{\mu\nu} \right) V_\nu = \beta_V \nabla^\mu \alpha. \label{NS_lim_Vmu}
\end{equation}
To obtain the constitutive relation for the diffusion current, we invert the tensor operator acting on $V_\nu$ on the left-hand side of the above equation. For this purpose, we employ the general decomposition of the diffusion current in a magnetic field~\cite{Panda:2020zhr},
\begin{equation} \label{diff_gen_sol}
    V_\nu = \left(\kappa_\parallel P^\parallel_{\nu\lambda} + \kappa_\perp P^\perp_{\nu\lambda} + \kappa_\times P^\times_{\nu\lambda}\right) \partial^\lambda \alpha.
\end{equation}
The tensors $P^\parallel_{\mu\nu}$, $P^\perp_{\mu\nu}$, and
$P^\times_{\mu\nu}$ are second-rank projection operators
that distinguish the different spatial components of the
diffusion current relative to the direction of the magnetic
field. The three transport coefficients, $\kappa_\parallel$,
$\kappa_\perp$, and $\kappa_\times$, characterize the
longitudinal, transverse, and Hall-like components of the
diffusion current, respectively. Their explicit forms are
obtained by inserting the tensor decomposition of $V_\nu$
into Eq.~(\ref{NS_lim_Vmu}) and solving the resulting tensor
equation, as discussed in Ref.~\cite{Panda:2020zhr}. With the quasiparticle description, we obtain
\begin{align}
    \kappa_{\parallel} &= \tau_c \beta_V,  \label{diff_coef1}
    \\
    \kappa_{\perp} &= \frac{\tau_c \beta_V}{1 + (q B \tau_c \delta_{VB})^2} , \label{diff_coef2}
    \\
    \kappa_{\times} &= \frac{\tau^2_c q B \beta_V \delta_{VB}}{1 + (q B \tau_c \delta_{VB})^2}. \label{diff_coef3}
\end{align}
\begin{figure}
    \centering
    \begin{minipage}[t]{0.47\linewidth}
        \includegraphics[width=\linewidth]{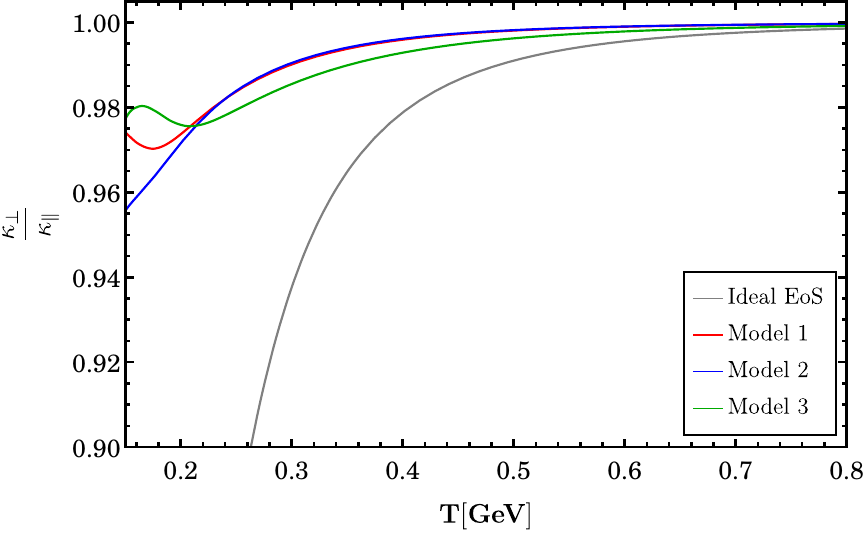}
    \end{minipage}
    \hfill
    \begin{minipage}[t]{0.46\linewidth}
        \includegraphics[width=\linewidth]{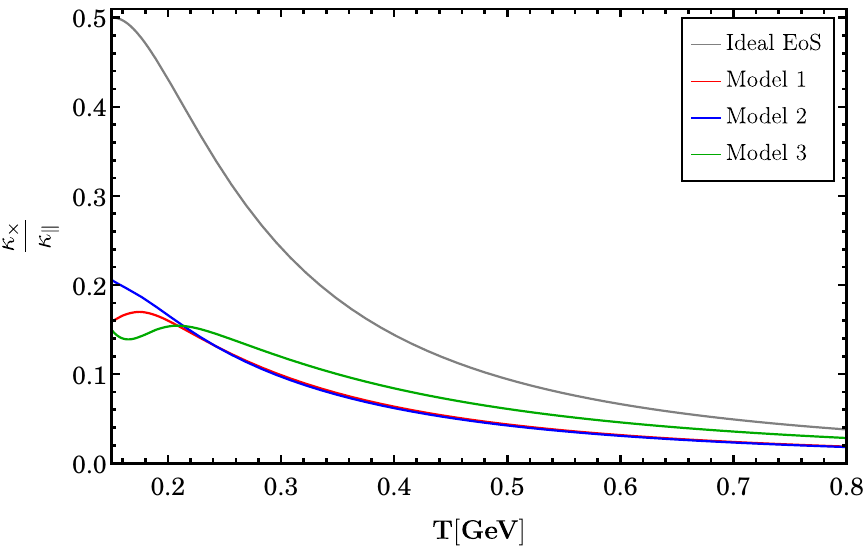}
    \end{minipage}
    \caption{Temperature dependence of magnetic field-induced anisotropic diffusion coefficients} 
    \label{kappa}
\end{figure}
The general solution for the diffusion current reflects the anisotropy introduced by an external magnetic field, giving rise to three distinct transport coefficients: the longitudinal, transverse, and Hall-like diffusion coefficients. Although the magnetic field is responsible for breaking the isotropy of the medium, interactions among the constituents of the QGP, incorporated through the quasiparticle framework, can also significantly influence these coefficients. The QCD EoS effects are embedded through $\beta_V$ and $\delta_{VB}$. To examine the effect of medium interactions on the magnetic field modified anisotropic diffusion coefficients, we investigate the temperature dependence of the ratios $\kappa_\perp/\kappa_\parallel$ and $\kappa_\perp/\kappa_\parallel$ within the effective description in Fig.~\ref{kappa}.  As the temperature increases, the
relative strength of the magnetic field decreases, leading to $\frac{\kappa_{\perp}}{\kappa_{\parallel}}
    \rightarrow 1$, $\frac{\kappa_{\times}}{\kappa_{\parallel}}
    \rightarrow 0$, recovering the expected isotropic behaviour of the medium.
These limiting trends indicate that the magnetic field-induced
anisotropy becomes progressively less significant at high
temperatures. This can be understood from the increasing
dominance of thermal motion over the effects associated with
the magnetic field. Fig.~\ref{kappa} also provides insight into the influence of the QCD medium interactions on magnetic-field-induced anisotropic diffusion. A comparison between the ideal and QCD EoS reveals that the QCD EoS suppresses the relative magnetic-field effects on diffusion, particularly in the low- and intermediate-temperature regions. This is reflected in the reduced deviation of $\kappa_\perp/\kappa_\parallel$ from unity and the smaller contribution of the Hall-like diffusion coefficient relative to the longitudinal one. The differences gradually diminish at higher temperatures, where both ratios approach their respective isotropic limits. These observations suggest that medium interactions, accounted for through the QCD EoS, play an important role in modifying the response of the QGP to an external magnetic field and consequently influence its anisotropic transport properties.
\section{Conclusion} \label{Conc}
In this work, we have investigated the second-order relativistic magnetohydrodynamic evolution of particle number diffusion in an interacting QCD medium. The medium interactions are incorporated through a temperature- and chemical potential-dependent effective mass, providing a thermodynamically consistent description of the QGP. We have solved the effective Boltzmann equation within RTA by employing the Chapman–Enskog-like iterative expansion method to obtain the non-equilibrium distribution function. The presence of the magnetic field modifies the tensorial structure of the diffusion dynamics and consequently alters the transport properties of the medium. We have provided a quantitative assessment of the interplay between the magnetic field and medium interactions using different parametrizations of the effective mass available in the literature.  Interestingly, our results reveal that the transport coefficients are strongly sensitive to the choice of effective-mass parametrization. We have also examined various limiting cases to establish the consistency of our analysis. In the Navier–Stokes limit, the magnetic field breaks the isotropy of the medium, leading to three distinct first-order transport coefficients associated with the longitudinal, transverse, and Hall-like components of the diffusion current. Our results show that the ideal EoS significantly overestimates the magnetic field-induced anisotropy in particle diffusion, particularly in the low and intermediate temperature regions. In contrast, the interacting QCD EoS suppresses the transverse and Hall-like diffusion contributions relative to the longitudinal component, indicating a weaker anisotropic response of the medium.

The present study highlights the importance of incorporating medium interactions in investigating dissipative transport in magnetized QCD matter. Further improvements to the present framework may be achieved by employing more realistic collision kernels beyond the RTA, which can provide a better description of microscopic interactions and their influence on transport coefficients. In particular, the extended RTA~\cite{Dash:2021ibx} and novel RTA~\cite{Rocha:2022fqz} could offer a more refined understanding of particle number diffusion in an interacting QCD medium under the influence of a magnetic field. However, extending the present second-order formulation to such generalized collision models, while maintaining thermodynamic consistency and properly accounting for the magnetic-field-dependent tensorial structures, remains a challenging task that deserves further investigation. Furthermore, at finite baryon density, the interplay between different conserved charges, including baryon number, electric charge, and strangeness, can give rise to cross-diffusion effects~\cite{Fotakis:2019nbq, Das:2021bkz, Sahu:2026vfc}. Investigating the corresponding coupled transport coefficients and their dependence on the magnetic field and medium interactions would be particularly relevant for the description of the QGP produced in low-energy heavy-ion collisions. We leave these aspects for future studies.

\acknowledgments
M.K. acknowledges the Department of Science and Technology (DST), Govt. of India, for the INSPIRE-Faculty award (DST/INSPIRE/04/2024/001794), and the Faculty Research Scheme (FRS project number: MISC 0240) at IIT (ISM) Dhanbad. 
\appendix

\section{Thermodynamic Integrals and Useful Relations} \label{Relations}

The $n$-th moment of thermodynamic integral is defined as:
\begin{equation}
    I^{(r)\pm}_{\mu_1 \mu_2 ... \mu_n} = \int \frac{dP_q}{(u \cdot p)^r} p_{\mu_1} p_{\mu_2}.....p_{\mu_n} (f_{q\,\,(0)}^+ \pm f_{q\,\,(0)}^-),
\end{equation}
which can be decomposed as,
\begin{equation}
    I^{(r)\pm}_{\mu_1 \mu_2 .... \mu_n} = I^{(r)\pm}_{n0} u_{\mu_1}...u_{\mu_n} + I^{(r)\pm}_{n1} (\Delta_{\mu_1 \mu_2} u_{\mu_3}...u_{\mu_n} + \text{perm.}) + ... + I^{(r)\pm}_{nq} (\Delta_{\mu_1 \mu_2} \Delta_{\mu_3 \mu_4} + \text{perm.}).
\end{equation}
The thermodynamic integral $I^{(r)\pm}_{nq}$ is defined as,
\begin{equation}
    I_{n q}^{(r)\pm} = \frac{(-1)^q}{(2q+1)!!} \int dP_q (u \cdot p)^{n-2q-r} (p \cdot \Delta \cdot p)^q  (f_{q\,\,(0)}^+ \pm f_{q\,\,(0)}^-).
\end{equation}
The integral representation of modified Bessel functions $K_n(z)$ and $K_{i,n}(z)$ are, 
\begin{equation}
    K_n(z) = \int_0^\infty d\theta \cosh(n\theta) \exp(-z \cosh\theta),
\end{equation}
\begin{equation}
    K_{i,n}(z) = \int_0^\infty d\theta \frac{1}{(\cosh\theta)^n} \exp(-z \cosh\theta).
\end{equation}
Recurrence relation of modified Bessel functions is as follows:
\begin{equation}
\begin{split}
    K_{n-1}(z) - K_{n+1}(z) = \frac{2 n}{z} K_n(z),
    \\
    K_{n-1}(z) + K_{n+1}(z) = -2 \ K'_n(z).
\end{split}
\end{equation}
The thermodynamic integral $I^{(r)\pm}_{n q}$ for both quarks and gluons can be written as: 
\begin{align}
    &I^{(r)+}_{nq} = (-1)^q \frac{g T^{n+2-r} z^{n+2-r}}{\pi^2 (2q+1)!!} \cosh{\alpha} \int_0^\infty d\theta (\cosh{\theta})^{n-2q-r} (\cosh^2{\theta} - 1)^{q+1} \exp(-z \cosh{\theta}),
    \\
    &I^{(r)-}_{nq} = (-1)^q \frac{g T^{n+2-r} z^{n+2-r}}{\pi^2 (2q+1)!!} \sinh{\alpha} \int_0^\infty d\theta (\cosh{\theta})^{n-2q-r} (\cosh^2{\theta} - 1)^{q+1} \exp(-z \cosh{\theta}),
    \\
    &I^{(r)}_{nq} = (-1)^q \frac{g T^{n+2-r} z^{n+2-r}}{2\pi^2 (2q+1)!!} \int_0^\infty d\theta (\cosh{\theta})^{n-2q-r} (\cosh^2{\theta} - 1)^{q+1} \exp(-z \cosh{\theta}).
\end{align}
For temperature and chemical potential-dependent effective mass $m(\alpha, \beta)$, its co-moving and spatial derivatives can be described as,
\begin{align}
    m\dot{m} &= m \left(\frac{\partial m}{\partial \beta} \dot{\beta} + \frac{\partial m}{\partial \alpha} \dot{\alpha} \right) = m \left(\frac{\partial m}{\partial \beta} \mathcal{X} + \frac{\partial m}{\partial \alpha} \mathcal{Y} \right)\theta,\\
     m \nabla^\mu m &= m \left(\frac{\partial m}{\partial \beta} \nabla^\mu \beta + \frac{\partial m}{\partial \alpha} \nabla^\mu \alpha \right), 
    \\
    &= -m \frac{\partial m}{\partial \beta} \left(\beta \dot{u}^\mu + \frac{\beta q B b^{\mu \lambda} V_\lambda}{\mathcal{E} + \mathcal{P}} \right) + m \left( \frac{n_f}{\mathcal{E} + \mathcal{P}} \frac{\partial m}{\partial \beta} + \frac{\partial m}{\partial \alpha} \right)\nabla^\mu \alpha.
\end{align}
\section{Second-order Diffusion Current Calculation}
Substituting the second-order correction to distribution function to the diffusion current, we obtain 
\begin{equation}
    V^\mu_2 = \Delta^\mu_\alpha \int dP_q \ p^\alpha \ (f_{q\,\,(2)}^+ - f_{q\,\,(2)}^-).
\end{equation}
Employing Eq.~(\ref{f_pm_2}), we define:

\begin{multline} \label{f_2_pm}
    f_{q\,\,(2)}^+ - f_{q\,\,(2)}^- = \underbrace{\frac{\tau_c}{u \cdot p} p^\gamma \partial_\gamma \left[\frac{\tau_c}{u \cdot p} p^\lambda \partial_\lambda(f_{q\,\,(0)}^+ - f_{q\,\,(0)}^-)\right]}_\text{1}
    + \underbrace{\frac{\tau_c}{u \cdot p} p^\gamma \partial_\gamma \left[\frac{\tau_c}{u \cdot p} m (\partial^\lambda m) \frac{\partial}{\partial p^\lambda}(f_{q\,\,(0)}^+ - f_{q\,\,(0)}^-) \right]}_\text{2} \\
    + \underbrace{\frac{\tau_c}{u \cdot p} m (\partial^\gamma m) \frac{\partial}{\partial p^\gamma} \left[\frac{\tau_c}{u \cdot p} p^\lambda \partial_\lambda (f_{q\,\,(0)}^+ - f_{q\,\,(0)}^-) \right]}_\text{3} 
    + \underbrace{\frac{\tau_c}{u \cdot p} m (\partial^\gamma m) \frac{\partial}{\partial p^\gamma} \left[ \frac{\tau_c}{u \cdot p} m (\partial^\lambda m) \frac{\partial}{\partial p^\lambda}(f_{q\,\,(0)}^+ - f_{q\,\,(0)}^-) \right]}_\text{4} \\
    - \underbrace{\frac{\tau_c}{u \cdot p} q B b^{\sigma\gamma} p_\gamma \frac{\partial}{\partial p^\sigma} \left[\frac{\tau_c}{u \cdot p} p^\lambda \partial_\lambda(f_{q\,\,(0)}^+ + f_{q\,\,(0)}^-) \right]}_\text{5}
    - \underbrace{\frac{\tau_c}{u \cdot p} q B b^{\sigma\gamma} p_\gamma \frac{\partial}{\partial p^\sigma} \left[\frac{\tau_c}{u \cdot p} m (\partial^\lambda m) \frac{\partial}{\partial p^\lambda}(f_{q\,\,(0)}^+ + f_{q\,\,(0)}^-) \right]}_\text{6} .
\end{multline}
Different terms appearing in Eq.~(\ref{f_2_pm}) are evaluated separately. Accordingly, the second-order diffusion current comprises as follows,
\begin{equation}
    V^\mu_2 = \mathcal{I}_1 + \mathcal{I}_2 + \mathcal{I}_3 + \mathcal{I}_4 + \mathcal{I}_5 + \mathcal{I}_6 ,
\end{equation}
where
\begin{align*}
    \mathcal{I}_1 &= \Delta^\mu_\alpha \int dP_q \ p^\alpha \frac{\tau_c}{u \cdot p} p^\gamma \partial_\gamma \left[\frac{\tau_c}{u \cdot p} p^\lambda \partial_\lambda(f_{q\,\,(0)}^+ - f_{q\,\,(0)}^-)\right],
    \\
    \mathcal{I}_2 &= \Delta^\mu_\alpha \int dP_q \ p^\alpha \frac{\tau_c}{u \cdot p} p^\gamma \partial_\gamma \left[\frac{\tau_c}{u \cdot p} m (\partial^\lambda m) \frac{\partial}{\partial p^\lambda}(f_{q\,\,(0)}^+ - f_{q\,\,(0)}^-) \right],
    \\
    \mathcal{I}_3 &= \Delta^\mu_\alpha \int dP_q \ p^\alpha \frac{\tau_c}{u \cdot p} m (\partial^\gamma m) \frac{\partial}{\partial p^\gamma} \left[\frac{\tau_c}{u \cdot p} p^\lambda \partial_\lambda (f_{q\,\,(0)}^+ - f_{q\,\,(0)}^-) \right],
    \\
    \mathcal{I}_4 &= \Delta^\mu_\alpha \int dP_q \ p^\alpha \frac{\tau_c}{u \cdot p} m (\partial^\gamma m) \frac{\partial}{\partial p^\gamma} \left[ \frac{\tau_c}{u \cdot p} m (\partial^\lambda m) \frac{\partial}{\partial p^\lambda}(f_{q\,\,(0)}^+ - f_{q\,\,(0)}^-) \right],
    \\
    \mathcal{I}_5 &= \Delta^\mu_\alpha \int dP_q \ p^\alpha \frac{\tau_c}{u \cdot p} q B b^{\sigma\gamma} p_\gamma \frac{\partial}{\partial p^\sigma} \left[\frac{\tau_c}{u \cdot p} p^\lambda \partial_\lambda(f_{q\,\,(0)}^+ + f_{q\,\,(0)}^-) \right],
    \\
    \mathcal{I}_6 &= \Delta^\mu_\alpha \int dP_q \ p^\alpha \frac{\tau_c}{u \cdot p} q B b^{\sigma\gamma} p_\gamma \frac{\partial}{\partial p^\sigma} \left[\frac{\tau_c}{u \cdot p} m (\partial^\lambda m) \frac{\partial}{\partial p^\lambda}(f_{q\,\,(0)}^+ + f_{q\,\,(0)}^-) \right].
\end{align*} 
We start with the first integral,
\begin{equation*}
    \mathcal{I}_1 = \Delta^{\mu}_{\alpha} \int dP_q \ p^\alpha \ \frac{\tau_c}{u \cdot p} p^\gamma \partial_\gamma \left[\frac{\tau_c}{u \cdot p} p^\lambda \partial_\lambda(f_{q\,\,(0)}^+ - f_{q\,\,(0)}^-)\right]=A_1+B_1+C_1,
\end{equation*}
where,
\begin{align*}
    A_1 &= \Delta^{\mu}_{\alpha} \int dP_q \ p^\alpha \tau_c D \left[\frac{\tau_c}{u \cdot p} p^\lambda \partial_\lambda(f_{q\,\,(0)}^+ - f_{q\,\,(0)}^-) \right],
    \\
    B_1 &= \Delta^{\mu}_{\alpha} \int dP_q \ p^\alpha \ \frac{\tau_c}{u \cdot p} p^\gamma \nabla_\gamma \left[\tau_c D(f_{q\,\,(0)}^+ - f_{q\,\,(0)}^-) \right],
    \\
    C_1 &= \Delta^{\mu}_{\alpha} \int dP_q \ p^\alpha \ \frac{\tau_c}{u \cdot p} p^\gamma \nabla_\gamma \left[\frac{\tau_c}{u \cdot p} p^\lambda \nabla_\lambda(f_{q\,\,(0)}^+ - f_{q\,\,(0)}^-) \right].
\end{align*}
Now, we evaluate each terms one by one. We obtain,
\begin{equation} \label{A1}
\begin{split}
    A_1 &= \Delta^{\mu}_{\alpha} \int dP_q \ p^\alpha \tau_c D \left[\frac{\tau_c}{u \cdot p} p^\lambda \partial_\lambda(f_{q\,\,(0)}^+ - f_{q\,\,(0)}^-) \right]
    \\
    &= -\Delta^{\mu}_{\alpha} D\left[ \int dP_q \ p^\alpha p^\lambda \ \frac{\tau^2_c}{u \cdot p} \Bigl\{(u \cdot p)\partial_\lambda \beta(f_{q\,\,(0)}^+ - f_{q\,\,(0)}^-) + \beta p^\sigma \partial_\lambda u_\sigma (f_{q\,\,(0)}^+ - f_{q\,\,(0)}^-) - \partial_\lambda \alpha (f_{q\,\,(0)}^+ + f_{q\,\,(0)}^-)\Bigr\} \right]
    \\
    &= -\tau_c \dot{V}^{\langle \mu \rangle} - \tau_c^2 \Delta^\mu_\alpha \ D \left[ \frac{n_f q B b^{\alpha\lambda} V_\lambda}{\mathcal{E} + \mathcal{P}} \right] - \tau_c^2 \left( I^{(0)-}_{20} \mathcal{X} - I^{(1)+}_{20} \mathcal{Y} + \beta I^{(0)-}_{21} \right) \dot{u}^\mu \theta ,
\end{split}
\end{equation}
\begin{equation}
\begin{split}
    B_1 &= \Delta^{\mu}_{\alpha} \int dP_q \ p^\alpha \ \frac{\tau_c}{u \cdot p} p^\gamma \nabla_\gamma \left[\tau_c D(f_{q\,\,(0)}^+ - f_{q\,\,(0)}^-) \right]
    \\
    &= \Delta^{\mu}_{\alpha} \nabla_\gamma \left[ \int dP_q \ p^\alpha p^\gamma \ \frac{\tau^2_c}{u \cdot p} D(f_{q\,\,(0)}^+ - f_{q\,\,(0)}^-) \right] + \Delta^{\mu}_{\alpha} \nabla_\gamma u_\sigma \left[\int dP_q \ p^\alpha p^\gamma p^\sigma \ \frac{\tau^2_c}{(u \cdot p)^2} D(f_{q\,\,(0)}^+ - f_{q\,\,(0)}^-) \right]
    \\
    &= -\tau^2_c \nabla^\mu \Bigl[I^{(0)-}_{21} \dot{\beta} - I^{(1)+}_{21} \dot{\alpha} \Bigr] - \tau^2_c \left[\frac{4}{3} I^{(0)-}_{21} + \frac{5}{3} I^{(2)-}_{42} \right] \beta \dot{u}^\mu \theta - \tau^2_c I^{(0)-}_{21} \beta \dot{u}_\lambda \omega^{\lambda\mu} - \tau^2_c \Bigl[I^{(0)-}_{21} + 2 I^{(2)-}_{42} \Bigr]  \beta \dot{u}_\lambda \sigma^{\lambda\mu} ,
\end{split}
\end{equation}
\begin{equation*}
\begin{split}
    C_1 &= \Delta^{\mu}_{\alpha} \int dP_q \ p^\alpha \ \frac{\tau_c}{u \cdot p} p^\gamma \nabla_\gamma \left[\frac{\tau_c}{u \cdot p} p^\lambda \nabla_\lambda(f_{q\,\,(0)}^+ - f_{q\,\,(0)}^-) \right]
    \\
    &= \Delta^{\mu}_{\alpha} \nabla_\gamma \left[ \int dP_q \ p^\alpha p^\gamma p^\lambda \ \frac{\tau^2_c}{(u \cdot p)^2} \nabla_\lambda(f_{q\,\,(0)}^+ - f_{q\,\,(0)}^-) \right] + \Delta^{\mu}_{\alpha} \nabla_\gamma u_\sigma \left[\int dP_q \ p^\alpha p^\sigma p^\gamma p^\lambda \ \frac{\tau^2_c}{(u \cdot p)^3} \nabla_\lambda(f_{q\,\,(0)}^+ - f_{q\,\,(0)}^-) \right] 
\end{split}
\end{equation*}
\begin{equation}
\begin{split}
     &= -\frac{4}{3} \tau^2_c \left[\frac{n_f}{\mathcal{E}+\mathcal{P}} I^{(0)-}_{21} - I^{(1)+}_{21}\right] (\nabla^\mu \alpha)\theta + \frac{4}{3} \tau^2_c I^{(0)-}_{21} \beta \dot{u}^\mu \theta - \tau^2_c \left[\frac{n_f}{\mathcal{E}+\mathcal{P}}I^{(0)-}_{21} - I^{(1)+}_{21}\right](\nabla^\gamma \alpha) \sigma^{\mu}_{\gamma} 
    \\
    &+ \tau^2_c \beta I^{(0)-}_{21} \dot{u}^\gamma \omega^\mu_\gamma - \tau^2_c \left[\frac{n_f}{\mathcal{E}+\mathcal{P}}I^{(0)-}_{21} - I^{(1)+}_{21}\right](\nabla^\gamma \alpha) \omega^{\mu}_{\gamma} +\tau^2_c \beta I^{(0)-}_{21} \dot{u}^\gamma \sigma^\mu_\gamma - 2\tau^2_c \Delta^\mu_\alpha \nabla_\gamma \Bigl[\beta I^{(2)-}_{42} \sigma^{\gamma\alpha} \Bigr]
    \\
    &-\frac{5}{3} \tau^2_c \nabla^\mu \Bigl[\beta I^{(2)-}_{42} \theta \Bigr] + \frac{5}{3} \tau^2_c I^{(2)-}_{42} \beta \dot{u}^\mu \theta + 2 \tau^2_c I^{(2)-}_{42} \beta \dot{u}^\lambda \sigma^\mu_\lambda - 2\tau^2_c \left[\frac{n_f}{\mathcal{E}+\mathcal{P}}I^{(2)-}_{42} - I^{(3)+}_{42} \right] (\nabla^\lambda \alpha) \sigma^\mu_\lambda
    \\
    &-\frac{5}{3} \tau^2_c  \left[\frac{n_f}{\mathcal{E}+\mathcal{P}}I^{(2)-}_{42} - I^{(3)+}_{42} \right] (\nabla^\mu \alpha) \theta + \frac{4}{3} \tau^2_c I^{(0)-}_{21} \theta \left[\frac{\beta q B b^{\mu \lambda} V_\lambda}{\mathcal{E} + \mathcal{P}} \right] + \tau^2_c I^{(0)-}_{21} \omega^\mu_\gamma \left[\frac{\beta q B b^{\gamma \lambda} V_\lambda}{\mathcal{E} + \mathcal{P}} \right]
    \\
    &+2 \tau^2_c I^{(2)-}_{42} \sigma^\mu_\gamma \left[\frac{\beta q B b^{\gamma \lambda} V_\lambda}{\mathcal{E} + \mathcal{P}} \right] +\tau^2_c I^{(0)-}_{21} \sigma^\mu_\gamma \left[\frac{\beta q B b^{\gamma \lambda} V_\lambda}{\mathcal{E} + \mathcal{P}} \right] + \frac{5}{3} \tau^2_c I^{(2)-}_{42} \theta \left[\frac{\beta q B b^{\mu \lambda} V_\lambda}{\mathcal{E} + \mathcal{P}} \right].
\end{split}
\end{equation}
The second integral can be written as,
\begin{equation*}
    \mathcal{I}_2 = \Delta^{\mu}_{\alpha} \int dP_q \ p^\alpha \ \frac{\tau_c}{u \cdot p} p^\gamma \partial_\gamma \left[\frac{\tau_c}{u \cdot p} m_ (\partial^\lambda m) \frac{\partial}{\partial p^\lambda}(f_{q\,\,(0)}^+ - f_{q\,\,(0)}^-) \right]=A_2+B_2+C_2.
\end{equation*}
The integral $\mathcal{I}_2$ can be decompose into three separate integrals as,
\begin{align*}
    A_2 &= -\Delta^{\mu}_{\alpha} \int dP_q \ p^\alpha p^\gamma \ \frac{\tau_c}{u \cdot p} \partial_\gamma \left (\frac{\tau_c}{u \cdot p} \right) \beta m\dot{m} \ (f_{q\,\,(0)}^+ - f_{q\,\,(0)}^-),
    \\
    B_2 &= -\Delta^{\mu}_{\alpha} \int dP_q \ p^\alpha p^\gamma \ \frac{\tau^2_c}{(u \cdot p)^2} \partial_\gamma (\beta m\dot{m}) \ (f_{q\,\,(0)}^+ - f_{q\,\,(0)}^-),
    \\
    C_2 &= -\Delta^{\mu}_{\alpha} \int dP_q \ p^\alpha p^\gamma \ \frac{\tau^2_c}{(u \cdot p)^2}
    \beta m\dot{m} \ \partial_\gamma(f_{q\,\,(0)}^+ - f_{q\,\,(0)}^-).
\end{align*}
Solving these integrals, we obtain,
\begin{align}
    A_2 &= -\Delta^{\mu}_{\alpha} \int dP_q \ p^\alpha p^\gamma \ \frac{\tau_c}{u \cdot p} \partial_\gamma \left (\frac{\tau_c}{u \cdot p} \right) \beta m\dot{m} \ (f_{q\,\,(0)}^+ - f_{q\,\,(0)}^-)
    = \tau^2_c I^{(2)-}_{21} \beta m \left(\frac{\partial m}{\partial \beta} \mathcal{X} + \frac{\partial m}{\partial \alpha} \mathcal{Y} \right) \dot{u}^\mu \theta ,\\
    B_2 &= -\Delta^{\mu}_{\alpha} \int dP_q \ p^\alpha p^\gamma \ \frac{\tau^2_c}{(u \cdot p)^2} \partial_\gamma (\beta m\dot{m}) \ (f_{q\,\,(0)}^+ - f_{q\,\,(0)}^-)
    =-\tau^2_c I^{(2)-}_{21} \nabla^\mu\left[\beta m \left(\frac{\partial m}{\partial \beta} \mathcal{X} + \frac{\partial m}{\partial \alpha} \mathcal{Y} \right) \theta \right],\\
    C_2 &= -\Delta^{\mu}_{\alpha} \int dP_q \ p^\alpha p^\gamma \ \frac{\tau^2_c}{(u \cdot p)^2} \beta m\dot{m} \ \partial_\gamma(f_{q\,\,(0)}^+ - f_{q\,\,(0)}^-) \nonumber,
    \\
    &= \tau^2_c \beta m \left( \frac{\partial m}{\partial \beta} \mathcal{X} + \frac{\partial m}{\partial \alpha} \mathcal{Y} \right) \left[ \frac{n_f}{\mathcal{E} + \mathcal{P}}I^{(1)-}_{21} - I^{(2)+}_{21} \right] (\nabla^\mu \alpha) \theta - \tau^2_c I^{(1)-}_{21}\beta m \left( \frac{\partial m}{\partial \beta} \mathcal{X} + \frac{\partial m}{\partial \alpha} \mathcal{Y} \right) \theta \frac{\beta q B b^{\mu\lambda} V_\lambda}{\mathcal{E} + \mathcal{P}}.
\end{align}
Third integral can be written as,
\begin{equation*}
    \mathcal{I}_3 = \Delta^{\mu}_{\alpha} \int dP_q \ p^\alpha \frac{\tau_c}{u \cdot p} m (\partial^\gamma m) \frac{\partial}{\partial p^\gamma} \left[\frac{\tau_c}{u \cdot p} p^\lambda \partial_\lambda (f_{q\,\,(0)}^+ - f_{q\,\,(0)}^-) \right]=A_3+B_3+C_3,
\end{equation*}
where,
\begin{align*}
    A_3 &= -\Delta^{\mu}_{\alpha} \int dP_q \ p^\alpha \frac{\tau_c}{u \cdot p} m (\partial^\gamma m) \frac{\partial}{\partial p^\gamma} \left[\tau_c p^\lambda (\partial_\lambda\beta) \ (f_{q\,\,(0)}^+ - f_{q\,\,(0)}^-) \right],
    \\
    B_3 &= -\Delta^{\mu}_{\alpha} \int dP_q \ p^\alpha \frac{\tau_c}{u \cdot p} m (\partial^\gamma m) \frac{\partial}{\partial p^\gamma} \left[\frac{\tau_c}{u \cdot p} \beta p^\lambda p^\sigma (\partial_\lambda u_\sigma) (f_{q\,\,(0)}^+ - f_{q\,\,(0)}^-)  \right],
    \\
    C_3 &= \Delta^{\mu}_{\alpha} \int dP_q \ p^\alpha \frac{\tau_c}{u \cdot p} m (\partial^\gamma m) \frac{\partial}{\partial p^\gamma} \left[\frac{\tau_c}{u \cdot p} p^\lambda (\partial_\lambda \alpha) (f_{q\,\,(0)}^+ + f_{q\,\,(0)}^-)  \right].
\end{align*}
Evaluating these integrals, we obtain
\begin{equation}
\begin{split}
    A_3 &= -\Delta^{\mu}_{\alpha} \int dP_q \ p^\alpha \frac{\tau_c}{u \cdot p} m (\partial^\gamma m) \frac{\partial}{\partial p^\gamma} \left[\tau_c p^\lambda (\partial_\lambda\beta) \ (f_{q\,\,(0)}^+ - f_{q\,\,(0)}^-) \right]
    \\
    &= -\tau^2_c I^{(1)-}_{21} \beta^2 m \left( \frac{\partial m}{\partial \beta} \mathcal{X} + \frac{\partial m}{\partial \alpha} \mathcal{Y} \right) \dot{u}^\mu \theta 
    + \tau^2_c I^{(1)-}_{21} \beta m \left( \frac{\partial m}{\partial \beta} \mathcal{X} + \frac{\partial m}{\partial \alpha} \mathcal{Y} \right) \frac{n_f}{\mathcal{E} + \mathcal{P}} (\nabla^\mu \alpha) \theta 
    \\
    &- \tau^2_c I^{(1)-}_{21} \beta m \left( \frac{\partial m}{\partial \beta} \mathcal{X} + \frac{\partial m}{\partial \alpha} \mathcal{Y} \right) \theta \frac{\beta q B b^{\mu \lambda} V_\lambda}{\mathcal{E} + \mathcal{P}},
\end{split}
\end{equation}
\begin{align}
    B_3 &= -\Delta^{\mu}_{\alpha} \int dP_q \ p^\alpha \frac{\tau_c}{u \cdot p} m (\partial^\gamma m) \frac{\partial}{\partial p^\gamma} \left[\frac{\tau_c}{u \cdot p} \beta p^\lambda p^\sigma (\partial_\lambda u_\sigma) (f_{q\,\,(0)}^+ - f_{q\,\,(0)}^-)  \right] \nonumber
    \\
    &=- \frac{2}{3} \beta m \tau^2_c I^{(2)-}_{21}  \theta \left[-\frac{\partial m}{\partial \beta} \left(\beta \dot{u}^\mu + \frac{\beta q B b^{\mu \gamma} V_\gamma}{\mathcal{E} + \mathcal{P}}\right) + \left( \frac{n_f}{\mathcal{E} + \mathcal{P}} \frac{\partial m}{\partial \beta} + \frac{\partial m}{\partial \alpha} \right) \nabla^\mu \alpha \right] \nonumber
    \\ 
    & - 2 \beta m \tau^2_c I^{(2)-}_{21} \sigma^\mu_\gamma \left[-\frac{\partial m}{\partial \beta} \left(\beta \dot{u}^\gamma + \frac{\beta q B b^{\gamma \rho} V_\rho}{\mathcal{E} + \mathcal{P}} \right) + \left( \frac{n_f}{\mathcal{E} + \mathcal{P}} \frac{\partial m}{\partial \beta} + \frac{\partial m}{\partial \alpha} \right) \nabla^\gamma \alpha \right]
    + \tau^2_c I^{(1)-}_{21} \beta^2 m \left( \frac{\partial m}{\partial \beta} \mathcal{X} + \frac{\partial m}{\partial \alpha} \mathcal{Y} \right) \dot{u}^\mu \theta,
    \\
    C_3 &= \Delta^{\mu}_{\alpha} \int dP_q \ p^\alpha \frac{\tau_c}{u \cdot p} m (\partial^\gamma m) \frac{\partial}{\partial p^\gamma} \left[\frac{\tau_c}{u \cdot p} p^\lambda (\partial_\lambda \alpha) (f_{q\,\,(0)}^+ + f_{q\,\,(0)}^-)  \right] \nonumber
    \\
    &= -\tau^2_c \Bigl(I^{(3)+}_{21} + I^{(2)+}_{21} \beta \Bigr) m \left( \frac{\partial m}{\partial \beta} \mathcal{X} + \frac{\partial m}{\partial \alpha} \mathcal{Y} \right) (\nabla^\mu \alpha)\theta.
\end{align}
Solving the fourth integral leads to,
\begin{equation}
    \mathcal{I}_4 = \Delta^\mu_\alpha \int dP_q \ p^\alpha \frac{\tau_c}{u \cdot p} m (\partial^\gamma m) \frac{\partial}{\partial p^\gamma} \left[ \frac{\tau_c}{u \cdot p} m (\partial^\lambda m) \frac{\partial}{\partial p^\lambda}(f_{q\,\,(0)}^+ - f_{q\,\,(0)}^-) \right] = 0.
\end{equation}
Solving the fifth integral, we obtain, 
\begin{equation}
\begin{split}
    \mathcal{I}_5 &= -\Delta^{\mu}_{\alpha} \int dP_q \ p^\alpha \frac{\tau_c}{u \cdot p} q B b^{\sigma\gamma} p_\gamma \frac{\partial}{\partial p^\sigma} \left[\frac{\tau_c}{u \cdot p} p^\lambda \partial_\lambda(f_{q\,\,(0)}^+ + f_{q\,\,(0)}^-) \right]
    \\
    &=-\tau^2_c\left[\frac{n_f}{\mathcal{E} + \mathcal{P}} I^{(1)+}_{21} - I^{(2)-}_{21} \right] q B b^{\mu \gamma} (\nabla_\gamma \alpha) 
    + \tau^2_c I^{(1)+}_{21} \beta q B b^{\gamma\mu}\left[\frac{\nabla_\gamma \Pi}{\mathcal{E} + \mathcal{P}}-\frac{\Pi \dot{u}_\gamma}{\mathcal{E} + \mathcal{P}} -\frac{\Delta^\sigma_\gamma \partial^\kappa \pi_{\kappa\sigma}}{\mathcal{E} + \mathcal{P}} \right].
\end{split}
\end{equation}
Solving the sixth integral leads to,
\begin{equation} \label{I2}
    \mathcal{I}_6 = -\Delta^\mu_\alpha \int dP_q \ p^\alpha \frac{\tau_c}{u \cdot p} q B b^{\sigma\gamma} p_\gamma \frac{\partial}{\partial p^\sigma} \left[\frac{\tau_c}{u \cdot p} m (\partial^\lambda m) \frac{\partial}{\partial p^\lambda}(f_{q\,\,(0)}^+ + f_{q\,\,(0)}^-) \right] = 0.
\end{equation}
Combining the Eqs.~(\ref{A1}) - (\ref{I2}), we get second-order evolution of the diffusion current as follows,
\begin{equation}
\begin{split}
    V^\mu_2 &= - \tau_c \dot{V}^{\langle \mu \rangle} - \tau_c^2 \Delta^\mu_\alpha D \left[\frac{n_f q B b^{\alpha \lambda} V_\lambda}{\mathcal{E} + \mathcal{P}} \right] - \tau_c^2 \left( I^{(0)-}_{20} \mathcal{X} - I^{(1)+}_{20} \mathcal{Y} + \beta I^{(0)-}_{21} \right) \dot{u}^\mu \theta 
    \\
    &-\tau_c^2 \nabla^\mu \left[\left(I^{(0)-}_{21} \mathcal{X} - I^{(1)+}_{21} \mathcal{Y} + \frac{5}{3} \beta I^{(2)-}_{42}\right) \theta\right] + 2\tau_c^2 \Delta^\mu_\alpha \nabla_\gamma \Bigl[\beta I^{(2)-}_{42} \sigma^{\gamma \alpha} \Bigr] - \tau_c^2 \left[\frac{n_f}{\mathcal{E} + \mathcal{P}} I^{(0)-}_{21} -I^{(1)+}_{21} \right] (\nabla^\gamma \alpha) \omega^\mu_\gamma 
    \\
    &-\tau_c^2 \left[\frac{4}{3} \left(\frac{n_f}{\mathcal{E} + \mathcal{P}} I^{(0)-}_{21} -I^{(1)+}_{21} \right) + \frac{5}{3}\left(\frac{n_f}{\mathcal{E} + \mathcal{P}} I^{(2)-}_{42} -I^{(3)+}_{42} \right) \right] (\nabla^\mu \alpha) \theta + \tau_c^2 \left[\frac{4}{3} I^{(0)-}_{21} + \frac{5}{3} I^{(2)-}_{42}\right] \theta \left[\frac{\beta q B b^{\mu\lambda} V_\lambda}{\mathcal{E} + \mathcal{P}} \right]
    \\
    &-\tau_c^2 \left[\left(\frac{n_f}{\mathcal{E} + \mathcal{P}} I^{(0)-}_{21} -I^{(1)+}_{21} \right) + 2 \left(\frac{n_f}{\mathcal{E} + \mathcal{P}} I^{(2)-}_{42} -I^{(3)+}_{42} \right) \right] (\nabla^\gamma \alpha) \sigma^\mu_\gamma + \tau_c^2 \left(I^{(0)-}_{21} + 2 I^{(2)-}_{42} \right) \sigma^\mu_\gamma \left[\frac{\beta q B b^{\gamma\lambda} V_\lambda}{\mathcal{E} + \mathcal{P}} \right] 
    \\
    &+ \tau_c^2 I^{(0)-}_{21} \omega^\mu_\gamma \left[\frac{\beta q B b^{\gamma\lambda} V_\lambda}{\mathcal{E} + \mathcal{P}} \right] + \tau_c^2 \left[\frac{n_f}{\mathcal{E} + \mathcal{P}} I^{(1)+}_{21} -I^{(2)-}_{21} \right] q B b^{\gamma\mu} (\nabla_\gamma \alpha) + \tau_c^2 I^{(1)+}_{21} \beta q B b^{\gamma\mu}\left[\frac{\nabla_\gamma \Pi}{\mathcal{E} + \mathcal{P}}-\frac{\Pi \dot{u}_\gamma}{\mathcal{E} + \mathcal{P}} -\frac{\Delta^\sigma_\gamma \partial^\kappa \pi_{\kappa\sigma}}{\mathcal{E} + \mathcal{P}} \right]
    \\
    &+ \tau_c^2 I^{(2)-}_{21} \beta m \left( \frac{\partial m}{\partial \beta} \mathcal{X} + \frac{\partial m}{\partial \alpha} \mathcal{Y} \right) \dot{u}^\mu \theta - \tau_c^2 I^{(2)-}_{21} \nabla^\mu \left[\beta m \left(\frac{\partial m}{\partial \beta} \mathcal{X} + \frac{\partial m}{\partial \alpha} \mathcal{Y} \right) \theta \right] - 2 \tau_c^2 I^{(1)-}_{21} \beta m \left( \frac{\partial m}{\partial \beta} \mathcal{X} + \frac{\partial m}{\partial \alpha} \mathcal{Y} \right) \theta \frac{\beta q B b^{\mu \lambda} V_\lambda}{\mathcal{E} + \mathcal{P}}
    \\
    &+ \tau_c^2 \beta m \left( \frac{\partial m}{\partial \beta} \mathcal{X} + \frac{\partial m}{\partial \alpha} \mathcal{Y} \right) \left[ \frac{n_f}{\mathcal{E} + \mathcal{P}}I^{(1)-}_{21} - I^{(2)+}_{21} \right] (\nabla^\mu \alpha) \theta - \tau_c^2 I^{(1)-}_{21}\beta m \left( \frac{\partial m}{\partial \beta} \mathcal{X} + \frac{\partial m}{\partial \alpha} \mathcal{Y} \right) \theta \frac{\beta q B b^{\mu\lambda} V_\lambda}{\mathcal{E} + \mathcal{P}}
    \\
    &+ \tau_c^2 I^{(1)-}_{21} \beta m \left( \frac{\partial m}{\partial \beta} \mathcal{X} + \frac{\partial m}{\partial \alpha} \mathcal{Y} \right) \frac{n_f}{\mathcal{E} + \mathcal{P}} (\nabla^\mu \alpha) \theta - \tau_c^2 \Bigl(I^{(3)+}_{21} + I^{(2)+}_{21} \beta \Bigr) m \left( \frac{\partial m}{\partial \beta} \mathcal{X} + \frac{\partial m}{\partial \alpha} \mathcal{Y} \right) (\nabla^\mu \alpha)\theta
    \\
    &+ 2 \tau_c^2 \beta^2 I^{(2)-}_{21} m \frac{\partial m}{\partial \beta} 
    \dot{u}^\lambda \sigma^\mu_\lambda - 2 \tau_c^2 \beta I^{(2)-}_{21} m \left( \frac{n_f}{\mathcal{E} + \mathcal{P}} \frac{\partial m}{\partial \beta} + \frac{\partial m}{\partial \alpha} \right) \sigma^\mu_\lambda (\nabla^\lambda \alpha) + 2 \tau_c^2 I^{(2)-}_{21} \beta m \frac{\partial m}{\partial \beta} \sigma^\mu_\lambda \frac{\beta q B b^{\lambda \rho} V_\rho}{\mathcal{E} + \mathcal{P}}
    \\
    &+ \frac{2}{3} \tau_c^2 \beta^2 I^{(2)-}_{21} m \frac{\partial m}{\partial \beta} \dot{u}^\mu \theta - \frac{2}{3} \tau_c^2 \beta I^{(2)-}_{21} m \left( \frac{n_f}{\mathcal{E} + \mathcal{P}} \frac{\partial m}{\partial \beta} + \frac{\partial m}{\partial \alpha} \right) (\nabla^\mu \alpha) \theta + \frac{2}{3} \tau_c^2 I^{(2)-}_{21} \beta m \frac{\partial m}{\partial \beta} \theta \frac{\beta q B b^{\mu \lambda} V_\lambda}{\mathcal{E} + \mathcal{P}}.
\end{split}
\end{equation}

\newpage
\section{Transport Coefficients Associated with Particle Diffusion } \label{Vmu_transp_coeff}

\renewcommand{\arraystretch}{1.5}
\setlength{\tabcolsep}{10pt}
\begin{table}[h] 
\begin{tabular}{|c|c|}
    \hline 
    $\lambda_{VV}$  & $\begin{array}{lcl}
        1 + \frac{2}{\beta_V} \left[\frac{n_f}{\mathcal{E} + \mathcal{P}} I^{(2)-}_{42} - I^{(3)+}_{42} \right] + \frac{2 \beta}{\beta_V} I^{(2)-}_{21} m \left[\frac{n_f}{\mathcal{E} + \mathcal{P}} \frac{\partial m}{\partial \beta}  + \frac{\partial m}{\partial \alpha}\right]
    \end{array}$  \\[1ex]
    \hline
    $\delta_{VV}$ & $\begin{array}{lcl} 
        &\frac{4}{3} + \frac{5}{3 \beta_V} \left[\frac{n_f}{\mathcal{E} + \mathcal{P}} I^{(2)-}_{42} - I^{(3)+}_{42} \right] - \frac{\beta}{\beta_V} m  \left(\frac{\partial m}{\partial \beta} \mathcal{X} + \frac{\partial m}{\partial \alpha} \mathcal{Y} \right) \left[ \frac{n_f}{\mathcal{E} + \mathcal{P}}I^{(1)-}_{21} - I^{(2)+}_{21} \right] \\ 
        &- \frac{\beta}{\beta_V} I^{(1)-}_{21}  m \left(\frac{\partial m}{\partial \beta} \mathcal{X} + \frac{\partial m}{\partial \alpha} \mathcal{Y} \right) \frac{n_f}{\mathcal{E} + \mathcal{P}}
    \end{array}$ \\ [1ex]
    \hline
    $l_{V\Pi}$ & $\begin{array}{lcl}
        -(A + I^{(0)-}_{21} B )= - \frac{1}{\beta_\Pi} \left[ I^{(2)-}_{21} \mathcal{X} - I^{(1)+}_{21} \mathcal{Y} + \frac{5}{3} \beta I^{(2)-}_{42} + \beta I^{(2)-}_{21} m \left(\frac{\partial m}{\partial \beta} \mathcal{X} + \frac{\partial m}{\partial \alpha} \mathcal{Y} \right) \right]
    \end{array}$ \\ [1ex]
    \hline
    $l_{V\pi}$ & $\begin{array}{lcl} 
        - \frac{\beta}{\beta_\pi} I^{(2)-}_{42}
    \end{array}$ \\ [1ex]
    \hline
    $\lambda_{V\pi}$ & $\begin{array}{lcl} 
        - \left[\frac{\partial l_{V\pi}}{\partial \beta} \frac{n_f}{\mathcal{E} + \mathcal{P}} + \frac{\partial l_{V\pi}}{\partial \alpha} \right] + \frac{\beta}{\beta_\pi} I^{(2)-}_{21} m \left[\frac{\partial m}{\partial \beta} \frac{n_f}{\mathcal{E} + \mathcal{P}} + \frac{\partial m}{\partial \alpha} \right]
    \end{array}$\\ [1ex]
    \hline
    $\tau_{V\pi}$ & $\begin{array}{lcl}
        \beta \frac{\partial l_{V\pi}}{\partial \beta} - l_{V\pi} - \beta I^{(2)-}_{21} m \frac{\partial m}{\partial \beta}
    \end{array}$ \\ [1ex]
    \hline
    $\tau_{V\Pi}$ & $\begin{array}{lcl}
         &-\beta \left(\frac{\partial A}{\partial \beta} + I^{(2)-}_{21} \frac{\partial B}{\partial \beta}  \right) + \left( I^{(0)-}_{20} \mathcal{X} - I^{(1)+}_{20} \mathcal{Y} + \beta I^{(0)-}_{21} \right) - \frac{\beta}{\beta_\Pi} I^{(2)-}_{21} m \left(\frac{\partial m}{\partial \beta} \mathcal{X} + \frac{\partial m}{\partial \alpha} \mathcal{Y} \right) 
    \\
    &- \frac{2\beta}{3\beta_\Pi} I^{(2)-}_{21} \beta m \frac{\partial m}{\partial \beta}
    \end{array}$\\ [1ex]
    \hline
    $\lambda_{V\Pi}$ & $\begin{array}{lcl}
        &\frac{n_f}{\mathcal{E} + \mathcal{P}} \left(\frac{\partial A}{\partial \beta} + I^{(2)-}_{21} \frac{\partial B}{\partial \beta} \right) + \left(\frac{\partial A}{\partial \alpha} + I^{(2)-}_{21} \frac{\partial B}{\partial \alpha} \right) + \frac{1}{\beta_\Pi} \Bigl(I^{(3)+}_{21} + \beta I^{(2)+}_{21} \Bigr) m \left(\frac{\partial m}{\partial \beta} \mathcal{X} + \frac{\partial m}{\partial \alpha} \mathcal{Y} \right)
        \\
        & + \frac{2\beta}{3\beta_\Pi} I^{(2)-}_{21} m \left[\frac{n_f}{\mathcal{E} + \mathcal{P}} \frac{\partial m}{\partial \beta} + \frac{\partial m}{\partial \alpha} \right] 
    \end{array}$ \\ [1ex]
    \hline
    $\delta_{VB}$    &  $\begin{array}{lcl}
        \frac{1}{\beta_V} \left[\frac{n_f}{\mathcal{E} + \mathcal{P}} I^{(1)+}_{21} - I^{(2)-}_{21} \right]
    \end{array}$ \\[1ex] 
    \hline
    $l_{V \pi B}$    & $\begin{array}{lcl}
        - \frac{\beta I^{(1)+}_{21}}{\mathcal{E} + \mathcal{P}}
    \end{array}$ \\[1ex]
    \hline
    $\tau_{V \Pi B}$ & $\begin{array}{lcl}
        - \frac{\beta I^{(1)+}_{21}}{\mathcal{E} + \mathcal{P}}
    \end{array}$ \\[1ex] 
    \hline
    $l_{V \Pi B}$    & $\begin{array}{lcl}
        - \frac{\beta I^{(1)+}_{21}}{\mathcal{E} + \mathcal{P}}
    \end{array}$ \\[1ex]
    \hline
    $\delta_{VVB}$   & $\begin{array}{lcl}
        & \left[\frac{\partial \tau_{VVB}}{\partial \beta} \mathcal{X} +\frac{\partial \tau_{VVB}}{\partial \alpha} \mathcal{Y} \right] - \frac{\beta}{\mathcal{E} + \mathcal{P}}\left[\frac{4}{3} I^{(0)-}_{21} + \frac{5}{3} I^{(2)-}_{42} \right] - \frac{\beta \beta_\Pi}{\mathcal{E} + \mathcal{P}} \left(\frac{\partial A}{\partial \beta} + I^{(2)-}_{21} \frac{\partial B}{\partial \beta} \right)
        \\
        &- \frac{2}{3} I^{(2)-}_{21} \frac{\beta^2}{\mathcal{E} + \mathcal{P}} m \frac{\partial m}{\partial \beta} + 2 I^{(1)-}_{21} \frac{\beta^2}{\mathcal{E} + \mathcal{P}} m \left(\frac{\partial m}{\partial \beta} \mathcal{X} + \frac{\partial m}{\partial \alpha} \mathcal{Y} \right)
    \end{array}$ \\[2ex] 
    \hline
    $\lambda_{VVB}$  & $\begin{array}{lcl}
        \frac{2 \beta \beta_\pi}{\mathcal{E} + \mathcal{P}} \frac{\partial l_{V\pi}}{\partial \beta} - \frac{\beta}{\mathcal{E} + \mathcal{P}} \left[I^{(0)-}_{21} + 2 I^{(2)-}_{42} \right] - 2 I^{(2)-}_{21} \frac{\beta^2}{\mathcal{E} + \mathcal{P}} m \frac{\partial m}{\partial \beta}
    \end{array}$ \\[1ex]
    \hline
    $\rho_{VVB}$     & $\begin{array}{lcl}
        - \frac{\beta I^{(0)-}_{21}}{\mathcal{E} + \mathcal{P}}
    \end{array}$ \\
    \hline
    $\tau_{VVB}$     & $\begin{array}{lcl}
         \frac{n_f}{\mathcal{E} + \mathcal{P}}
    \end{array}$ \\ 
    \hline
\end{tabular}
\end{table}
Here, 
\begin{align*}
    A &= \frac{1}{\beta_\Pi} \left(I^{(0)-}_{21} \mathcal{X} - I^{(1)+}_{21} \mathcal{Y} + \frac{5}{3} \beta I^{(2)-}_{42}\right),
    \\
    B &= \frac{\beta}{\beta_\Pi} m \left(\frac{\partial m}{\partial \beta} \mathcal{X} + \frac{\partial m}{\partial \beta} \mathcal{Y}\right) .
\end{align*}

\bibliography{new_ref}

@article{Mukherjee:2025dqp,
    author = "Mukherjee, Arghya and Bhadury, Samapan and Singha, Pracheta",
    title = "{Quasiparticle hydrodynamics with momentum-dependent relaxation time}",
    eprint = "2504.11572",
    archivePrefix = "arXiv",
    primaryClass = "hep-ph",
    doi = "10.1103/nkx2-bxys",
    journal = "Phys. Rev. D",
    volume = "112",
    number = "5",
    pages = "056018",
    year = "2025"
}

@article{0608:Bannur,
    author = "Bannur, Vishnu M.",
    title = "{Comments on quasiparticle models of quark-gluon plasma}",
    eprint = "hep-ph/0608232",
    archivePrefix = "arXiv",
    doi = "10.1016/j.physletb.2007.02.030",
    journal = "Phys. Lett. B",
    volume = "647",
    pages = "271--274",
    year = "2007"
}

@article{Panda:2020zhr,
    author = "Panda, Ankit Kumar and Dash, Ashutosh and Biswas, Rajesh and Roy, Victor",
    title = "{Relativistic non-resistive viscous magnetohydrodynamics from the kinetic theory: a relaxation time approach}",
    eprint = "2011.01606",
    archivePrefix = "arXiv",
    primaryClass = "nucl-th",
    doi = "10.1007/JHEP03(2021)216",
    journal = "JHEP",
    volume = "03",
    pages = "216",
    year = "2021"
}

@article{Daher:2024vxk,
    author = "Daher, Asaad and Tinti, Leonardo and Jaiswal, Amaresh and Ryblewski, Radoslaw",
    title = "{Quasiparticle second-order dissipative hydrodynamics at finite chemical potential}",
    eprint = "2412.06024",
    archivePrefix = "arXiv",
    primaryClass = "hep-ph",
    doi = "10.1103/PhysRevD.111.074011",
    journal = "Phys. Rev. D",
    volume = "111",
    number = "7",
    pages = "074011",
    year = "2025"
}

@article{Anderson1974466,
    author = "Anderson, J. L. and Witting, H. R.",
    title = "{A relativistic relaxation-time model for the Boltzmann equation}",
    doi = "10.1016/0031-8914(74)90355-3",
    journal = "Physica",
    volume = "74",
    number = "3",
    pages = "466--488",
    year = "1974"
}

@article{Tinti:2016bav,
    author = "Tinti, Leonardo and Jaiswal, Amaresh and Ryblewski, Radoslaw",
    title = "{Quasiparticle second-order viscous hydrodynamics from kinetic theory}",
    eprint = "1612.07329",
    archivePrefix = "arXiv",
    primaryClass = "nucl-th",
    doi = "10.1103/PhysRevD.95.054007",
    journal = "Phys. Rev. D",
    volume = "95",
    number = "5",
    pages = "054007",
    year = "2017"
}

@article{Rocha:2024rce,
    author = "Rocha, Gabriel S. and Denicol, Gabriel S.",
    title = "{Transport coefficients of transient hydrodynamics for the hadron-resonance gas and thermal-mass quasiparticle models}",
    eprint = "2402.06996",
    archivePrefix = "arXiv",
    primaryClass = "nucl-th",
    doi = "10.1103/PhysRevD.109.096011",
    journal = "Phys. Rev. D",
    volume = "109",
    number = "9",
    pages = "096011",
    year = "2024"
}

@article{Borsanyi:2010cj,
    author = "Borsanyi, Szabolcs and Endrodi, Gergely and Fodor, Zoltan and Jakovac, Antal and Katz, Sandor D. and Krieg, Stefan and Ratti, Claudia and Szabo, Kalman K.",
    title = "{The QCD equation of state with dynamical quarks}",
    eprint = "1007.2580",
    archivePrefix = "arXiv",
    primaryClass = "hep-lat",
    reportNumber = "WUB-10-12",
    doi = "10.1007/JHEP11(2010)077",
    journal = "JHEP",
    volume = "11",
    pages = "077",
    year = "2010"
}

@article{Bluhm:2007nu,
    author = "Bluhm, M. and Kampfer, Burkhard and Schulze, R. and Seipt, D. and Heinz, U.",
    title = "{A family of equations of state based on lattice QCD: Impact on flow in ultrarelativistic heavy-ion collisions}",
    eprint = "0705.0397",
    archivePrefix = "arXiv",
    primaryClass = "hep-ph",
    doi = "10.1103/PhysRevC.76.034901",
    journal = "Phys. Rev. C",
    volume = "76",
    pages = "034901",
    year = "2007"
}

@article{Gale:2013da,
    author = "Gale, Charles and Jeon, Sangyong and Schenke, Bjoern",
    title = "{Hydrodynamic Modeling of Heavy-Ion Collisions}",
    eprint = "1301.5893",
    archivePrefix = "arXiv",
    primaryClass = "nucl-th",
    doi = "10.1142/S0217751X13400113",
    journal = "Int. J. Mod. Phys. A",
    volume = "28",
    pages = "1340011",
    year = "2013"
}

@article{Heinz:2013th,
    author = "Heinz, Ulrich and Snellings, Raimond",
    title = "{Collective flow and viscosity in relativistic heavy-ion collisions}",
    eprint = "1301.2826",
    archivePrefix = "arXiv",
    primaryClass = "nucl-th",
    doi = "10.1146/annurev-nucl-102212-170540",
    journal = "Ann. Rev. Nucl. Part. Sci.",
    volume = "63",
    pages = "123--151",
    year = "2013"
}

@article{Kurian:2018qwb,
    author = "Kurian, Manu and Mitra, Sukanya and Ghosh, Snigdha and Chandra, Vinod",
    title = "{Transport coefficients of hot magnetized QCD matter beyond the lowest Landau level approximation}",
    eprint = "1805.07313",
    archivePrefix = "arXiv",
    primaryClass = "nucl-th",
    doi = "10.1140/epjc/s10052-019-6649-z",
    journal = "Eur. Phys. J. C",
    volume = "79",
    number = "2",
    pages = "134",
    year = "2019"
}

@article{Inghirami:2019mkc,
    author = "Inghirami, Gabriele and Mace, Mark and Hirono, Yuji and Del Zanna, Luca and Kharzeev, Dmitri E. and Bleicher, Marcus",
    title = "{Magnetic fields in heavy ion collisions: flow and charge transport}",
    eprint = "1908.07605",
    archivePrefix = "arXiv",
    primaryClass = "hep-ph",
    doi = "10.1140/epjc/s10052-020-7847-4",
    journal = "Eur. Phys. J. C",
    volume = "80",
    number = "3",
    pages = "293",
    year = "2020"
}

@article{Fukushima:2008xe,
    author = "Fukushima, Kenji and Kharzeev, Dmitri E. and Warringa, Harmen J.",
    title = "{The Chiral Magnetic Effect}",
    eprint = "0808.3382",
    archivePrefix = "arXiv",
    primaryClass = "hep-ph",
    doi = "10.1103/PhysRevD.78.074033",
    journal = "Phys. Rev. D",
    volume = "78",
    pages = "074033",
    year = "2008"
}

@article{Hattori:2016emy,
    author = "Hattori, Koichi and Huang, Xu-Guang",
    title = "{Novel quantum phenomena induced by strong magnetic fields in heavy-ion collisions}",
    eprint = "1609.00747",
    archivePrefix = "arXiv",
    primaryClass = "nucl-th",
    reportNumber = "RBRC-1202",
    doi = "10.1007/s41365-016-0178-3",
    journal = "Nucl. Sci. Tech.",
    volume = "28",
    number = "2",
    pages = "26",
    year = "2017"
}

@article{Kharzeev:2015znc,
    author = "Kharzeev, D. E. and Liao, J. and Voloshin, S. A. and Wang, G.",
    title = "{Chiral magnetic and vortical effects in high-energy nuclear collisions{\textemdash}A status report}",
    eprint = "1511.04050",
    archivePrefix = "arXiv",
    primaryClass = "hep-ph",
    doi = "10.1016/j.ppnp.2016.01.001",
    journal = "Prog. Part. Nucl. Phys.",
    volume = "88",
    pages = "1--28",
    year = "2016"
}

@article{Li:2017tgi,
    author = "Li, Shiyong and Yee, Ho-Ung",
    title = "{Shear viscosity of the quark-gluon plasma in a weak magnetic field in perturbative QCD: Leading log}",
    eprint = "1707.00795",
    archivePrefix = "arXiv",
    primaryClass = "hep-ph",
    doi = "10.1103/PhysRevD.97.056024",
    journal = "Phys. Rev. D",
    volume = "97",
    number = "5",
    pages = "056024",
    year = "2018"
}

@article{Ghosh:2023ghi,
    author = "Ghosh, Ritesh and Jamal, Mohammad Yousuf and Kurian, Manu",
    title = "{Impact of the chiral asymmetry and a magnetic field on the passage of an energetic test parton in a QCD medium}",
    eprint = "2306.10247",
    archivePrefix = "arXiv",
    primaryClass = "hep-ph",
    doi = "10.1103/PhysRevD.108.054035",
    journal = "Phys. Rev. D",
    volume = "108",
    number = "5",
    pages = "054035",
    year = "2023"
}

@article{ALICE:2019sgg,
    author = "Acharya, Shreyasi and others",
    collaboration = "ALICE",
    title = "{Probing the effects of strong electromagnetic fields with charge-dependent directed flow in Pb-Pb collisions at the LHC}",
    eprint = "1910.14406",
    archivePrefix = "arXiv",
    primaryClass = "nucl-ex",
    reportNumber = "CERN-EP-2019-241",
    doi = "10.1103/PhysRevLett.125.022301",
    journal = "Phys. Rev. Lett.",
    volume = "125",
    number = "2",
    pages = "022301",
    year = "2020"
}

@article{Jiang:2022uoe,
    author = "Jiang, Ze-Fang and Cao, Shanshan and Xing, Wen-Jing and Wu, Xiang-Yu and Yang, C. B. and Zhang, Ben-Wei",
    title = "{Probing the initial longitudinal density profile and electromagnetic field in ultrarelativistic heavy-ion collisions with heavy quarks}",
    eprint = "2202.13555",
    archivePrefix = "arXiv",
    primaryClass = "nucl-th",
    doi = "10.1103/PhysRevC.105.054907",
    journal = "Phys. Rev. C",
    volume = "105",
    number = "5",
    pages = "054907",
    year = "2022"
}

@article{Rath:2025zhw,
    author = "Rath, Shubhalaxmi and Neill, Nicol{\'a}s A.",
    title = "{Studying the thermoelectric properties of an anisotropic QGP medium}",
    eprint = "2511.03054",
    archivePrefix = "arXiv",
    primaryClass = "hep-ph",
    doi = "10.1103/rjkc-9gdc",
    journal = "Phys. Rev. D",
    volume = "113",
    number = "7",
    pages = "076004",
    year = "2026"
}

@article{Das:2016cwd,
    author = "Das, Santosh K. and Plumari, Salvatore and Chatterjee, Sandeep and Alam, Jane and Scardina, Francesco and Greco, Vincenzo",
    title = "{Directed Flow of Charm Quarks as a Witness of the Initial Strong Magnetic Field in Ultra-Relativistic Heavy Ion Collisions}",
    eprint = "1608.02231",
    archivePrefix = "arXiv",
    primaryClass = "nucl-th",
    doi = "10.1016/j.physletb.2017.02.046",
    journal = "Phys. Lett. B",
    volume = "768",
    pages = "260--264",
    year = "2017"
}

@article{STAR:2019clv,
    author = "Adam, Jaroslav and others",
    collaboration = "STAR",
    title = "{First Observation of the Directed Flow of $D^{0}$ and $\overline{D^0}$ in Au+Au Collisions at $\sqrt{s_{\rm NN}}$ = 200 GeV}",
    eprint = "1905.02052",
    archivePrefix = "arXiv",
    primaryClass = "nucl-ex",
    doi = "10.1103/PhysRevLett.123.162301",
    journal = "Phys. Rev. Lett.",
    volume = "123",
    number = "16",
    pages = "162301",
    year = "2019"
}

@article{Karmakar:2019tdp,
    author = "Karmakar, Bithika and Ghosh, Ritesh and Bandyopadhyay, Aritra and Haque, Najmul and Mustafa, Munshi G.",
    title = "{Anisotropic pressure of deconfined QCD matter in presence of strong magnetic field within one-loop approximation}",
    eprint = "1902.02607",
    archivePrefix = "arXiv",
    primaryClass = "hep-ph",
    doi = "10.1103/PhysRevD.99.094002",
    journal = "Phys. Rev. D",
    volume = "99",
    number = "9",
    pages = "094002",
    year = "2019"
}

@article{Kurian:2017yxj,
    author = "Kurian, Manu and Chandra, Vinod",
    title = "{Effective description of hot QCD medium in strong magnetic field and longitudinal conductivity}",
    eprint = "1709.08320",
    archivePrefix = "arXiv",
    primaryClass = "nucl-th",
    doi = "10.1103/PhysRevD.96.114026",
    journal = "Phys. Rev. D",
    volume = "96",
    number = "11",
    pages = "114026",
    year = "2017"
}

@article{Ghosh:2024fkg,
    author = "Ghosh, Ritesh and Shovkovy, Igor A.",
    title = "{Electrical conductivity of hot relativistic plasma in a strong magnetic field}",
    eprint = "2404.01388",
    archivePrefix = "arXiv",
    primaryClass = "hep-ph",
    doi = "10.1103/PhysRevD.110.096009",
    journal = "Phys. Rev. D",
    volume = "110",
    number = "9",
    pages = "096009",
    year = "2024"
}

@article{Ghosh:2022vjp,
    author = "Ghosh, Ritesh and Kurian, Manu",
    title = "{Magnetic-field-dependent electric-charge transport in hadronic medium at finite temperature}",
    eprint = "2211.06729",
    archivePrefix = "arXiv",
    primaryClass = "hep-ph",
    doi = "10.1103/PhysRevC.107.034903",
    journal = "Phys. Rev. C",
    volume = "107",
    number = "3",
    pages = "034903",
    year = "2023"
}

@article{Dash:2020vxk,
    author = "Dash, Ashutosh and Samanta, Subhasis and Dey, Jayanta and Gangopadhyaya, Utsab and Ghosh, Sabyasachi and Roy, Victor",
    title = "{Anisotropic transport properties of a hadron resonance gas in a magnetic field}",
    eprint = "2002.08781",
    archivePrefix = "arXiv",
    primaryClass = "nucl-th",
    doi = "10.1103/PhysRevD.102.016016",
    journal = "Phys. Rev. D",
    volume = "102",
    number = "1",
    pages = "016016",
    year = "2020"
}

@article{Wang:2022jxx,
    author = "Wang, Xinyang and Shovkovy, Igor A.",
    title = "{Rate and ellipticity of dilepton production in a magnetized quark-gluon plasma}",
    eprint = "2205.00276",
    archivePrefix = "arXiv",
    primaryClass = "nucl-th",
    doi = "10.1103/PhysRevD.106.036014",
    journal = "Phys. Rev. D",
    volume = "106",
    number = "3",
    pages = "036014",
    year = "2022"
}

@article{Satapathy:2021cjp,
    author = "Satapathy, Sarthak and Ghosh, Snigdha and Ghosh, Sabyasachi",
    title = "{Kubo estimation of the electrical conductivity for a hot relativistic fluid in the presence of a magnetic field}",
    eprint = "2104.03917",
    archivePrefix = "arXiv",
    primaryClass = "hep-ph",
    doi = "10.1103/PhysRevD.104.056030",
    journal = "Phys. Rev. D",
    volume = "104",
    number = "5",
    pages = "056030",
    year = "2021"
}

@article{Denicol:2019iyh,
    author = "Denicol, Gabriel S. and Moln\'ar, Etele and Niemi, Harri and Rischke, Dirk H.",
    title = "{Resistive dissipative magnetohydrodynamics from the Boltzmann-Vlasov equation}",
    eprint = "1902.01699",
    archivePrefix = "arXiv",
    primaryClass = "nucl-th",
    doi = "10.1103/PhysRevD.99.056017",
    journal = "Phys. Rev. D",
    volume = "99",
    number = "5",
    pages = "056017",
    year = "2019"
}

@article{Denicol:2018rbw,
    author = "Denicol, Gabriel S. and Huang, Xu-Guang and Moln\'ar, Etele and Monteiro, Gustavo M. and Niemi, Harri and Noronha, Jorge and Rischke, Dirk H. and Wang, Qun",
    title = "{Nonresistive dissipative magnetohydrodynamics from the Boltzmann equation in the 14-moment approximation}",
    eprint = "1804.05210",
    archivePrefix = "arXiv",
    primaryClass = "nucl-th",
    doi = "10.1103/PhysRevD.98.076009",
    journal = "Phys. Rev. D",
    volume = "98",
    number = "7",
    pages = "076009",
    year = "2018"
}

@article{Gangopadhyaya:2022olm,
    author = "Gangopadhyaya, Utsab and Roy, Victor",
    title = "{Order-by-order anisotropic transport coefficients of a magnetised fluid: a Chapman-Enskog approach}",
    eprint = "2206.12197",
    archivePrefix = "arXiv",
    primaryClass = "nucl-th",
    doi = "10.1007/JHEP09(2022)114",
    journal = "JHEP",
    volume = "09",
    pages = "114",
    year = "2022"
}

@article{Kurian:2019nna,
    author = "Kurian, Manu and Das, Santosh K. and Chandra, Vinod",
    title = "{Heavy quark dynamics in a hot magnetized QCD medium}",
    eprint = "1907.09556",
    archivePrefix = "arXiv",
    primaryClass = "nucl-th",
    doi = "10.1103/PhysRevD.100.074003",
    journal = "Phys. Rev. D",
    volume = "100",
    number = "7",
    pages = "074003",
    year = "2019"
}

@article{Panda:2021pvq,
    author = "Panda, Ankit Kumar and Dash, Ashutosh and Biswas, Rajesh and Roy, Victor",
    title = "{Relativistic resistive dissipative magnetohydrodynamics from the relaxation time approximation}",
    eprint = "2104.12179",
    archivePrefix = "arXiv",
    primaryClass = "nucl-th",
    doi = "10.1103/PhysRevD.104.054004",
    journal = "Phys. Rev. D",
    volume = "104",
    number = "5",
    pages = "054004",
    year = "2021"
}

@article{Bzdak:2011yy,
    author = "Bzdak, Adam and Skokov, Vladimir",
    title = "{Event-by-event fluctuations of magnetic and electric fields in heavy ion collisions}",
    eprint = "1111.1949",
    archivePrefix = "arXiv",
    primaryClass = "hep-ph",
    reportNumber = "BNL-96541-2011-JA, RBRC-927",
    doi = "10.1016/j.physletb.2012.02.065",
    journal = "Phys. Lett. B",
    volume = "710",
    pages = "171--174",
    year = "2012"
}

@article{Deng:2012pc,
    author = "Deng, Wei-Tian and Huang, Xu-Guang",
    title = "{Event-by-event generation of electromagnetic fields in heavy-ion collisions}",
    eprint = "1201.5108",
    archivePrefix = "arXiv",
    primaryClass = "nucl-th",
    doi = "10.1103/PhysRevC.85.044907",
    journal = "Phys. Rev. C",
    volume = "85",
    pages = "044907",
    year = "2012"
}

@article{Tuchin:2013ie,
    author = "Tuchin, Kirill",
    title = "{Particle production in strong electromagnetic fields in relativistic heavy-ion collisions}",
    eprint = "1301.0099",
    archivePrefix = "arXiv",
    primaryClass = "hep-ph",
    doi = "10.1155/2013/490495",
    journal = "Adv. High Energy Phys.",
    volume = "2013",
    pages = "490495",
    year = "2013"
}

@article{Romatschke:2007mq,
    author = "Romatschke, Paul and Romatschke, Ulrike",
    title = "{Viscosity Information from Relativistic Nuclear Collisions: How Perfect is the Fluid Observed at RHIC?}",
    eprint = "0706.1522",
    archivePrefix = "arXiv",
    primaryClass = "nucl-th",
    reportNumber = "INT-PUB-07-14",
    doi = "10.1103/PhysRevLett.99.172301",
    journal = "Phys. Rev. Lett.",
    volume = "99",
    pages = "172301",
    year = "2007"
}

@article{Schenke:2010rr,
    author = "Schenke, Bjorn and Jeon, Sangyong and Gale, Charles",
    title = "{Elliptic and triangular flow in event-by-event (3+1)D viscous hydrodynamics}",
    eprint = "1009.3244",
    archivePrefix = "arXiv",
    primaryClass = "hep-ph",
    doi = "10.1103/PhysRevLett.106.042301",
    journal = "Phys. Rev. Lett.",
    volume = "106",
    pages = "042301",
    year = "2011"
}

@article{DenicolTau,
    author = "Denicol, G. S. and Jeon, S. and Gale, C.",
    title = "{Transport Coefficients of Bulk Viscous Pressure in the 14-moment approximation}",
    eprint = "1403.0962",
    archivePrefix = "arXiv",
    primaryClass = "nucl-th",
    doi = "10.1103/PhysRevC.90.024912",
    journal = "Phys. Rev. C",
    volume = "90",
    number = "2",
    pages = "024912",
    year = "2014"
}

@article{Singh:2024leo,
    author = "Singh, Sunny Kumar and Kurian, Manu and Chandra, Vinod",
    title = "{Revisiting shear stress tensor evolution: Nonresistive magnetohydrodynamics with momentum-dependent relaxation time}",
    eprint = "2403.13160",
    archivePrefix = "arXiv",
    primaryClass = "hep-ph",
    doi = "10.1103/PhysRevD.110.014004",
    journal = "Phys. Rev. D",
    volume = "110",
    number = "1",
    pages = "014004",
    year = "2024"
}

@article{Florkowski:2015lra,
    author = "Florkowski, Wojciech and Jaiswal, Amaresh and Maksymiuk, Ewa and Ryblewski, Radoslaw and Strickland, Michael",
    title = "{Relativistic quantum transport coefficients for second-order viscous hydrodynamics}",
    eprint = "1503.03226",
    archivePrefix = "arXiv",
    primaryClass = "nucl-th",
    doi = "10.1103/PhysRevC.91.054907",
    journal = "Phys. Rev. C",
    volume = "91",
    pages = "054907",
    year = "2015"
}

@article{Bhadury:2020ivo,
    author = "Bhadury, Samapan and Kurian, Manu and Chandra, Vinod and Jaiswal, Amaresh",
    title = "{Second order relativistic viscous hydrodynamics within an effective description of hot QCD medium}",
    eprint = "2010.01537",
    archivePrefix = "arXiv",
    primaryClass = "hep-ph",
    doi = "10.1088/1361-6471/ac1d9a",
    journal = "J. Phys. G",
    volume = "48",
    number = "10",
    pages = "105104",
    year = "2021"
}

@article{Hattori:2017qih,
    author = "Hattori, Koichi and Huang, Xu-Guang and Rischke, Dirk H. and Satow, Daisuke",
    title = "{Bulk Viscosity of Quark-Gluon Plasma in Strong Magnetic Fields}",
    eprint = "1708.00515",
    archivePrefix = "arXiv",
    primaryClass = "hep-ph",
    doi = "10.1103/PhysRevD.96.094009",
    journal = "Phys. Rev. D",
    volume = "96",
    number = "9",
    pages = "094009",
    year = "2017"
}

@article{Hattori:2022hyo,
    author = "Hattori, Koichi and Hongo, Masaru and Huang, Xu-Guang",
    title = "{New Developments in Relativistic Magnetohydrodynamics}",
    eprint = "2207.12794",
    archivePrefix = "arXiv",
    primaryClass = "hep-th",
    reportNumber = "RIKEN-iTHEMS-Report-22",
    doi = "10.3390/sym14091851",
    journal = "Symmetry",
    volume = "14",
    number = "9",
    pages = "1851",
    year = "2022"
}

@article{Hattori:2020htm,
    author = "Hattori, Koichi and Taya, Hidetoshi and Yoshida, Shinsuke",
    title = "{Di-lepton production from a single photon in strong magnetic fields: vacuum dichroism}",
    eprint = "2010.13492",
    archivePrefix = "arXiv",
    primaryClass = "hep-ph",
    reportNumber = "YITP-20-136",
    doi = "10.1007/JHEP01(2021)093",
    journal = "JHEP",
    volume = "01",
    pages = "093",
    year = "2021"
}

@article{Chandra:2011en,
    author = "Chandra, Vinod and Ravishankar, V.",
    title = "{A quasi-particle description of $(2+1)$- flavor lattice QCD equation of state}",
    eprint = "1103.0091",
    archivePrefix = "arXiv",
    primaryClass = "nucl-th",
    reportNumber = "TIFR-TH-11-06",
    doi = "10.1103/PhysRevD.84.074013",
    journal = "Phys. Rev. D",
    volume = "84",
    pages = "074013",
    year = "2011"
}

@article{Rocha:2022fqz,
    author = "Rocha, Gabriel S. and Ferreira, Maur{\'\i}cio N. and Denicol, Gabriel S. and Noronha, Jorge",
    title = "{Transport coefficients of quasiparticle models within a new relaxation time approximation of the Boltzmann equation}",
    eprint = "2203.15571",
    archivePrefix = "arXiv",
    primaryClass = "nucl-th",
    doi = "10.1103/PhysRevD.106.036022",
    journal = "Phys. Rev. D",
    volume = "106",
    number = "3",
    pages = "036022",
    year = "2022"
}

@article{Das:2019ppb,
    author = "Das, Arpan and Mishra, Hiranmaya and Mohapatra, Ranjita K.",
    title = "{Electrical conductivity and Hall conductivity of a hot and dense quark gluon plasma in a magnetic field: A quasiparticle approach}",
    eprint = "1907.05298",
    archivePrefix = "arXiv",
    primaryClass = "hep-ph",
    doi = "10.1103/PhysRevD.101.034027",
    journal = "Phys. Rev. D",
    volume = "101",
    number = "3",
    pages = "034027",
    year = "2020"
}

@article{Dash:2021ibx,
    author = "Dash, Dipika and Bhadury, Samapan and Jaiswal, Sunil and Jaiswal, Amaresh",
    title = "{Extended relaxation time approximation and relativistic dissipative hydrodynamics}",
    eprint = "2112.14581",
    archivePrefix = "arXiv",
    primaryClass = "nucl-th",
    doi = "10.1016/j.physletb.2022.137202",
    journal = "Phys. Lett. B",
    volume = "831",
    pages = "137202",
    year = "2022"
}

@article{Modak:2026pmi,
    author = "Modak, Ashes and Shaikh, Anowar and Kurian, Manu and Panda, Binata and Dash, Sadhana",
    title = "{Electromagnetic response of a relativistic drifting plasma}",
    eprint = "2604.27800",
    archivePrefix = "arXiv",
    primaryClass = "hep-ph",
    doi = "10.1103/sl21-xj4r",
    journal = "Phys. Rev. D",
    volume = "114",
    number = "3",
    pages = "034048",
    year = "2026"
}

@article{Das:2021bkz,
    author = "Das, Arpan and Mishra, Hiranmaya and Mohapatra, Ranjita K.",
    title = "{Diffusion matrix associated with the diffusion processes of multiple conserved charges in a hot and dense hadronic matter}",
    eprint = "2109.01543",
    archivePrefix = "arXiv",
    primaryClass = "nucl-th",
    doi = "10.1103/PhysRevD.106.014013",
    journal = "Phys. Rev. D",
    volume = "106",
    number = "1",
    pages = "014013",
    year = "2022"
}

@article{Sahu:2026vfc,
    author = "Sahu, Dushmanta and Goswami, Kangkan and Sahoo, Raghunath",
    title = "{Conserved-charge diffusion in hot and dense QCD matter}",
    eprint = "2610.16829",
    archivePrefix = "arXiv",
    primaryClass = "hep-ph",
    month = "10",
    year = "2026"
}

@article{Fotakis:2019nbq,
    author = "Fotakis, Jan A. and Greif, Moritz and Greiner, Carsten and Denicol, Gabriel S. and Niemi, Harri",
    title = "{Diffusion processes involving multiple conserved charges: A study from kinetic theory and implications to the fluid-dynamical modeling of heavy ion collisions}",
    eprint = "1912.09103",
    archivePrefix = "arXiv",
    primaryClass = "hep-ph",
    doi = "10.1103/PhysRevD.101.076007",
    journal = "Phys. Rev. D",
    volume = "101",
    number = "7",
    pages = "076007",
    year = "2020"
}

@article{Goswami:2023eol,
    author = "Goswami, Kangkan and Sahu, Dushmanta and Dey, Jayanta and Sahoo, Raghunath and Stock, Reinhard",
    title = "{Anisotropy of magnetized quark matter}",
    eprint = "2310.02711",
    archivePrefix = "arXiv",
    primaryClass = "hep-ph",
    doi = "10.1103/PhysRevD.109.074012",
    journal = "Phys. Rev. D",
    volume = "109",
    number = "7",
    pages = "074012",
    year = "2024"
}

@article{Rath:2023abm,
    author = "Rath, Shubhalaxmi and Dash, Sadhana",
    title = "{Nonextensive effects on the viscous properties of hot and magnetized QCD matter}",
    eprint = "2307.12002",
    archivePrefix = "arXiv",
    primaryClass = "hep-ph",
    doi = "10.1140/epja/s10050-024-01252-3",
    journal = "Eur. Phys. J. A",
    volume = "60",
    number = "2",
    pages = "29",
    year = "2024"
}

@article{Shaikh:2022sky,
    author = "Shaikh, Anowar and Rath, Shubhalaxmi and Dash, Sadhana and Panda, Binata",
    title = "{Flow of charge and heat in thermal QCD within the weak magnetic field limit: A Bhatnagar-Gross-Krook model approach}",
    eprint = "2210.15388",
    archivePrefix = "arXiv",
    primaryClass = "hep-ph",
    doi = "10.1103/PhysRevD.108.056021",
    journal = "Phys. Rev. D",
    volume = "108",
    number = "5",
    pages = "056021",
    year = "2023"
}

@article{Bluhm:2006yh,
    author = "Bluhm, Marcus and Kampfer, Burkhard and Schulze, Robert and Seipt, Daniel",
    editor = "Antinori, F. and Bass, S. and De Falco, A. and Kuhn, C. and Nardi, M. and Peitzmann, T. and Ullrich, T. and Velkovska, J. and Wiedemann, U. A.",
    title = "{Quasi-Particle Description of Strongly Interacting Matter: Towards a Foundation}",
    eprint = "hep-ph/0608053",
    archivePrefix = "arXiv",
    doi = "10.1140/epjc/s10052-006-0056-y",
    journal = "Eur. Phys. J. C",
    volume = "49",
    pages = "205--211",
    year = "2007"
}

@article{Mykhaylova:2019wci,
    author = "Mykhaylova, Valeriya and Bluhm, Marcus and Redlich, Krzysztof and Sasaki, Chihiro",
    title = "{Quark-flavor dependence of the shear viscosity in a quasiparticle model}",
    eprint = "1906.01697",
    archivePrefix = "arXiv",
    primaryClass = "hep-ph",
    doi = "10.1103/PhysRevD.100.034002",
    journal = "Phys. Rev. D",
    volume = "100",
    number = "3",
    pages = "034002",
    year = "2019"
}

@article{Plumari:2011mk,
    author = "Plumari, Salvatore and Alberico, Wanda M. and Greco, Vincenzo and Ratti, Claudia",
    title = "{Recent thermodynamic results from lattice QCD analyzed within a quasi-particle model}",
    eprint = "1103.5611",
    archivePrefix = "arXiv",
    primaryClass = "hep-ph",
    doi = "10.1103/PhysRevD.84.094004",
    journal = "Phys. Rev. D",
    volume = "84",
    pages = "094004",
    year = "2011"
}

@article{Berrehrah:2016vzw,
    author = "Berrehrah, Hamza and Bratkovskaya, Elena and Steinert, Thorsten and Cassing, Wolfgang",
    title = "{A dynamical quasiparticle approach for the QGP bulk and transport properties}",
    eprint = "1605.02371",
    archivePrefix = "arXiv",
    primaryClass = "hep-ph",
    doi = "10.1142/S0218301316420039",
    journal = "Int. J. Mod. Phys. E",
    volume = "25",
    number = "07",
    pages = "1642003",
    year = "2016"
}

@article{Bannur:2006js,
    author = "Bannur, Vishnu M.",
    title = "{Self-consistent quasiparticle model for quark-gluon plasma}",
    eprint = "hep-ph/0609188",
    archivePrefix = "arXiv",
    doi = "10.1103/PhysRevC.75.044905",
    journal = "Phys. Rev. C",
    volume = "75",
    pages = "044905",
    year = "2007"
}

@article{Shi:2019wzi,
    author = "Shi, Shuzhe and Zhang, Hui and Hou, Defu and Liao, Jinfeng",
    title = "{Signatures of Chiral Magnetic Effect in the Collisions of Isobars}",
    eprint = "1910.14010",
    archivePrefix = "arXiv",
    primaryClass = "nucl-th",
    doi = "10.1103/PhysRevLett.125.242301",
    journal = "Phys. Rev. Lett.",
    volume = "125",
    pages = "242301",
    year = "2020"
}

@article{Panda:2025yxw,
    author = "Panda, Ankit Kumar and Das, Aritra and Dash, Ashutosh and Bandyopadhyay, Aritra and Islam, Chowdhury Aminul",
    title = "{Dilepton Spectra and Even Flow Harmonics in a Magnetized QGP: An Ideal Hydrodynamic Study}",
    eprint = "2508.15035",
    archivePrefix = "arXiv",
    primaryClass = "hep-ph",
    month = "8",
    year = "2025"
}

@article{Dash:2022xkz,
    author = "Dash, Ashutosh and Shokri, Masoud and Rezzolla, Luciano and Rischke, Dirk H.",
    title = "{Charge diffusion in relativistic resistive second-order dissipative magnetohydrodynamics}",
    eprint = "2211.09459",
    archivePrefix = "arXiv",
    primaryClass = "nucl-th",
    doi = "10.1103/PhysRevD.107.056003",
    journal = "Phys. Rev. D",
    volume = "107",
    number = "5",
    pages = "056003",
    year = "2023"
}
\end{document}